\documentclass[11pt,a4paper]{article}
\usepackage{epsf,amssymb,graphicx,scalefnt}
\usepackage[tbtags]{amsmath}
\usepackage{cite}

\newcommand{\abbrev}{\rm\scalefont{.9}}
\newcommand{\lhc}{{\abbrev LHC}}
\newcommand{\sm}{{\abbrev SM}}
\newcommand{\mssm}{{\abbrev MSSM}}
\newcommand{\susy}{{\abbrev SUSY}}
\newcommand{\lsp}{{\abbrev LSP}}
\newcommand{\brpv}{{\abbrev BRpV}}
\newcommand{\cp}{{\abbrev CP}}
\newcommand{\lo}{{\abbrev LO}}
\newcommand{\nlo}{{\abbrev NLO}}
\newcommand{\cmb}{{\abbrev CMB}}
\newcommand{\pmns}{{\abbrev PMNS}}

\newcommand{\drbar}{\overline{\mbox{\abbrev DR}}}
\newcommand{\eqn}[1]{Eq.\,(\ref{#1})}

\newcommand{\fig}[1]{Fig.\,\ref{#1}}
\newcommand{\tab}[1]{Tab.\,\ref{#1}}
\newcommand{\sct}[1]{Sect.\,\ref{#1}}
\newcommand{\citere}[1]{Ref.~\cite{#1}}
\newcommand{\citeres}[1]{Refs.~\cite{#1}}
\newcommand{\Nge}{\mathcal{N}}
\newcommand{\Mass}{\mathcal{M}}
\newcommand{\hS}{\hat{\Sigma}} 
\newcommand{\sla}[1]{\ooalign{\hfil\hfil/\hfil\crcr$#1$}} 
\newcommand{\Lag}{\mathcal{L}}
\newcommand{\prog}{{\tt CNNDecays}}
\newcommand{\chiN}{\tilde{\chi}_1^0}
\newcommand{\Br}{\mathrm{Br}}
\newcommand{\tN}{\widetilde{N}}

\date{}
\title{\vspace*{-2em}
  \begin{flushright}
    {\sf\small DESY 14-109}
  \end{flushright}
  \vspace*{2em} \cp{}~violation in bilinear $R$-parity violation\\
  and its consequences for the early universe}
  \author{Asma Ch\'erigu\`ene$^a$, Stefan Liebler$^b$, Werner Porod$^a$\\[2em] 
  $^a$ {\it Institut f\"ur Theoretische Physik und Astrophysik, Universit\"at W\"urzburg}\\
  {\it 97074  W\"urzburg, Germany}\\[1em]
  $^b$ {\it II. Institut f\"ur Theoretische Physik, Universit\"at Hamburg}\\
  {\it 22761 Hamburg, Germany}}
    
\begin{document}
\maketitle

\begin{abstract}
Supersymmetric models with bilinear $R$-parity violation (\brpv) provide a framework for neutrino masses and
mixing angles to explain neutrino oscillation data. We consider \cp{}~violation within the new physical
phases in \brpv{} and discuss their effect on the generation of neutrino masses and
the decays of the lightest supersymmetric particle (\lsp{}), being a light neutralino with mass $\sim 100$\,GeV,
at next-to-leading order. 
The decays affect the lepton and via sphaleron transitions the baryon asymmetry in the early universe.
For a rather light \lsp{}, asymmetries generated before the electroweak
phase transition via e.g. the Affleck-Dine mechanism are reduced up to
two orders of magnitude, but are still present. On the other hand, the decays of a light \lsp{}
themselves can account for the generation of a lepton and baryon asymmetry,
the latter in accordance to the observation in our universe, since the smallness
of the \brpv{} parameters allows for an out-of-equilibrium decay and sufficiently
large \cp{}~violation is possible consistent with experimental bounds from the non-observation
of electric dipole-moments.
\end{abstract}

\pagebreak

\section{Introduction}
\label{sec:introduction}

The observed baryonic component of the universe comes along with
the question why the universe consists of entirely matter
with hardly any primordial antimatter \cite{Beringer:1900zz}.
Defining the baryon and antibaryon
number density $n_B$ and $n_{\bar{B}}$ and the
entropy~$s$ at temperature~$T$, the
baryon asymmetry can be expressed in terms of the quantity
\begin{equation}
\delta_B = \frac{n_B-n_{\bar{B}}}{s} \quad.
\end{equation}
The value of $\delta_B$, which is consistent
with the primordial abundances of light elements originating
from big bang nucleosynthesis \cite{Copi:1994ev,kolbbook}, is
$\delta_B=(6.19 \pm 0.14) \cdot 10^{-10}$
extracted from the measurements of the acoustic peaks in the cosmic microwave
background (\cmb{}) \cite{Bennett:2003bz,Bennett:2012zja}.

The dynamical creation of the baryon asymmetry in the universe (Baryogenesis)
requires the implementation of the three Sakharov conditions \cite{Sakharov:1967dj}:
violation of baryon number~$B$, {\abbrev C} symmetry and \cp{}~symmetry violation
and  departure from thermal equilibrium.
Nonperturbative effects (sphalerons) \cite{'tHooft:1976fv}
can give rise to processes, which
conserve $B-L$ with $L$ being the lepton number
of involved particles, but violate $B+L$. Thus, a generated lepton
asymmetry can account for the observed baryon asymmetry as well
(Baryogenesis via Leptogenesis \cite{Fukugita:1986hr,Khlebnikov:1988sr,Harvey:1990qw}),
in particular since lepton asymmetries are hardly constrained by
experiments \cite{Serpico:2005bc,Simha:2008mt,Popa:2008tb}.

In bilinear $R$-parity violation, where $L$ violating
parameters allow for the generation of neutrino masses and
mixing, the decays of the lightest supersymmetric particle (\lsp{})
can thus affect the lepton and baryon asymmetries in the
early universe after inflation. Whereas $L$ violation is explicitly
given by the \brpv{} parameters, we incorporate \cp{}~violation
by complex phases for those parameters.
Lastly, the \lsp{} decay widths are small enough
to be out of equilibrium, if the \brpv{} parameters are chosen in agreement
with neutrino masses and mixing and the \lsp{} is rather light, e.g.\
$m_{\tilde \chi^0_1} \lesssim 100$~GeV.
We study the evolution of the number densities by solving numerically
the corresponding Boltzmann equations. On the one hand
existing asymmetries
e.g.\ induced by the Affleck-Dine mechanism \cite{Affleck:1984fy}
are reduced by up to two orders of magnitude, but are still present.
On the other hand we demonstrate that \cp{}~violating \lsp{} decays
can generate lepton asymmetries. Before the electroweak phase transition
the latter asymmetries can be partially transferred to baryon asymmetries
via sphaleron transitions in accordance to the observation.

Earlier works on Leptogenesis in the context of \brpv{}
\cite{Hambye:1999pw,Hambye:2000zs,Hambye:2001eu,BenDayan:2005gu,Chakrabortty:2011zz}
made use of complex gaugino masses, leaving the
$R$-parity violating parameters real. In those cases only small lepton asymmetries
below $10^{-10}$ can be generated, if the \lsp{} is supposed to decay
out of equilibrium. However, nonholomorphic terms in combination with fixed
particle spectra \cite{Hambye:2000zs} induce small enough decay widths for the neutralino
to be out of equilibrium and allow for large enough \cp{}~asymmetries in
the decay products being a charged Higgs boson and a lepton.
As it was pointed in \citere{Hambye:1999pw} lepton number
violating decays can also spoil existing lepton asymmetries.

We focus on \brpv{} parameters, which are in agreement
with the observations of neutrino oscillations.
In accordance to the global fit carried out
in \citeres{Tortola:2012te,GonzalezGarcia:2012sz,Capozzi:2013csa,Forero:2014bxa} 
the preferred and in our analysis employed ranges
of the oscillation parameters at $2\sigma$ (for a normal neutrino mass hierarchy) are given by \citere{Capozzi:2013csa}
\footnote{Similar values are found by the most recent global fit in \citere{Forero:2014bxa}.}
\begin{align}\nonumber
 0.376 \leq &\sin^2\theta_{23} \leq 0.506~,\qquad 2.30 \times10^{-3}~\mathrm{eV}^2 \leq \Delta m_{31}^2 \leq 2.59 \times10^{-3}~\mathrm{eV}^2\\\nonumber
 0.275 \leq &\sin^2\theta_{12} \leq 0.342~,\qquad 7.15 \times10^{-5}~\mathrm{eV}^2 \leq \Delta m_{21}^2 \leq 8.00 \times10^{-5}~\mathrm{eV}^2\\
 0.0197 \leq &\sin^2\theta_{13} \leq 0.0276\quad.
\end{align}
The explanation of neutrino masses and mixing within \brpv{} was widely
discussed in the literature, for reviews we refer to \citeres{Barbier:2004ez,Chemtob:2004xr,Hirsch:2004he}.
In \brpv{} it is well-known that lowest order in perturbation
theory is not sufficient to generate the full neutrino spectrum, however loop corrections
can nicely explain the mass hierarchy between the neutrino mass eigenstates
\cite{Hempfling:1995wj,Roy:1996bua,Diaz:1997xc,Bisset:1998bt,
Joshipura:1998fn,Hirsch:2000ef,Abada:2001zh,Joshipura:2002fc,
Chun:2002vp,Diaz:2003as,Dedes:2006ni}.
We shortly repeat the discussion, but focus mainly on complex phases in
the \brpv{} parameters, which additionally induce a Dirac \cp{}~phase
in the lepton/neutrino mixing matrix. \cp{}~violation in the partial decay widths
of the \lsp{} occurs at the one-loop level \cite{Hirsch:2002tq}.
Our calculation of \lsp{} decays at next-to-leading order (\nlo{})
is based on \citeres{Liebler:2010bi,Liebler:2011tp}.

The remainder of this paper is organized as follows:
In \sct{sec:brpvcp} we explain the theory behind \brpv{} for complex \brpv{} parameters.
This discussion includes the generation of neutrino masses and the calculation
of the \lsp{} decay at \nlo{} in the electromagnetic coupling.
Moreover we provide the basics of number
density evolution in the early universe by the introduction of Boltzmann equations.
Afterwards we shortly present a simple description of the transition between a lepton
and baryon asymmetry via sphaleron transitions.
In \sct{sec:results} we show our numerical results starting again with neutrino
masses and mixing and the neutralino decays. In case of \cp{}~conserving \brpv{}
initial asymmetries can be reduced, being
up to two orders of magnitude lesser in size.
For \cp{}~violation instead
the neutralino decays themselves provide a large lepton asymmetry, which is partially
transformed to a baryon asymmetry due to sphaleron transitions.
In the last subsection we elaborate on the effects of \lsp{} annihilation to \sm{}
particles in more detail. Finally we conclude in \sct{sec:conclusions} and
present the implemented Boltzmann equations in the Appendix.

\section{Bilinear $R$-parity violation and \cp{}~violation therein}
\label{sec:brpvcp}

For bilinear $R$-parity violation (\brpv{}), which was first discussed
in \citeres{Hall:1983id,Lee:1984kr,Lee:1984tn,Ross:1984yg,Ellis:1984gi}, 
the superpotential is given by
\begin{flalign}
\label{eq:supW}
 W \, =  \, \varepsilon_{ab}[Y^{ij}_U\hat{Q}^a_i\hat{U}^c_j\hat{H}^b_u
 +Y^{ij}_D\hat{Q}^b_i\hat{D}^c_j\hat{H}^a_d
 +Y^{ij}_E\hat{L}^b_i\hat{E}^c_j\hat{H}^a_d
 -\mu \hat{H}^a_d\hat{H}^b_u+\epsilon_i\hat{L}^a_i\hat{H}^b_u]\quad,
\end{flalign}
where $Y_U, Y_D$ and $Y_E$ are the $(3\times3)$ Yukawa matrices and 
$\varepsilon_{\alpha\beta}$ is the complete antisymmetric SU$(2)$ tensor with $\varepsilon_{12}=1$,
whereas $i,j$ denote the three generations of leptons and quarks.
The last terms $\epsilon_i$ explicitly break lepton number~$L$ and are similar to the parameter~$\mu$,
which determines the mass of the Higgsinos, given in units of mass.
Additionally the three soft-\susy{} breaking parameters $B_i$ are added to
the minimal supersymmetric standard model (\mssm{}) soft-breaking Lagrangian
\begin{flalign}
\begin{split}
 \label{eq:softsusy}
&\mathcal{L}_{soft}= M^{ij2}_Q\tilde{Q}^{a*}_i\tilde{Q}^{a}_j+M^{ij2}_U\tilde{U}_i\tilde{U}^*_j+ 
M^{ij2}_D\tilde{D}_i\tilde{D}^*_j+M^{ij2}_L\tilde{L}^{a*}_i\tilde{L}^a_j+ 
M^{ij2}_E\tilde{E}_i\tilde{E}^*_j\\
&\quad +m^2_{H_d}H^{a*}_d H^a_d+m^2_{H_u}H^{a*}_u H^a_u-
\frac{1}{2}\Bigl[M_1\tilde{B}^0\tilde{B}^0 +M_2 \tilde{W}^\gamma\tilde{W}^\gamma+M_3\tilde{g}^{\gamma'}\tilde{g}^{\gamma'}+h.c.\Bigr]\\
&\quad +\varepsilon_{ab}\Bigl[T^{ij}_U\tilde{Q}^a_i\tilde{U}_j^*{H}^b_u+T^{ij}_D\tilde{Q}^b_i\hat{D}_j^*{H}^a_d+
T^{ij}_E\tilde{L}^b_i\tilde{E}_j^*H_d^a - B_\mu \mu H^a_d H^b_u -B_i \epsilon_i \tilde{L}^a_i H^b_u+h.c.\Bigr],
\end{split} 
\end{flalign}
where a summation over $a,b\in\lbrace 1,2\rbrace$, $\gamma\in\lbrace 1,2,3\rbrace$
and $\gamma'\in\lbrace 1,\ldots,8\rbrace$ and the
generation indices $i$ and $j$ is implied.
The vacuum structure induces
vacuum expectation values (VEVs) for the
neutral components of the Higgs fields $\langle H_u^0\rangle = v_u/\sqrt{2}$
and $\langle H_d^0\rangle = v_d/\sqrt{2}$ as well as the 
sneutrinos $\langle \tilde{\nu}_i \rangle = v_i/\sqrt{2}$.
The latter VEVs together with the last term in \eqn{eq:supW}
result in a mixing between the gauge eigenstates of the neutralinos 
$\tilde{B},~\tilde{W}^0_3,~\tilde{H}^0_d$ and $\tilde{H}^0_u$ and the three left-handed neutrinos 
$\nu_i$ at tree-level, providing an effective Majorana mass term for the neutrinos at tree-level
\cite{Nowakowski:1995dx,Hirsch:2000ef,Diaz:2003as}.
Moreover the charginos mix with the charged leptons and the scalars, pseudoscalars and charged scalar
states have to be combined with the sneutrinos and sleptons respectively.

To study the effects of \cp{}~violation in \brpv{} we closely follow
\citere{Hirsch:2002tq} and allow for complex parameters
\begin{flalign}
\epsilon_i=\epsilon^R_i+i\epsilon^I_i,\quad B_i=B_i^R+iB_i^I,\quad
B_\mu=B_\mu^R+iB_\mu^I
\label{eq:complexpara}
\end{flalign}
in the superpotential \eqn{eq:supW} and the soft-breaking terms in \eqn{eq:softsusy}.
Our phase convention is such that the gaugino mass parameter $M_2$
is real and positive.
To simplify our model, $\mu$ and all other parameters in the soft-breaking Lagrangian are
taken to be real, although additional complex phases are possible. In this way, the
constraints on the electric dipole-moments of electron, neutron and various atoms are
satisfied if the parameters are chosen to fulfill neutrino data \cite{Hirsch:2002tq}. In case of \cp{}~violation
scalars and pseudoscalars are indistinguishable. The resulting mass matrix 
is shown in \citere{Hirsch:2002tq}. We choose the VEVs $v_d, v_u$ and $v_i$ to be real
and determine the real and complex parts of $B_\mu$ and $B_i$ from the tadpole equations.
Due to the real $\mu$ parameter it yields $B_\mu^I\propto \sum_i v_i\epsilon_i^I$ and $B_i^I\propto \epsilon_i^I$,
such that the real \brpv{} model is restored in the limit $\epsilon_i^I\rightarrow 0$.

\subsection{Neutrino masses and mixing angles}
\label{sec:neutrinomassestree}

In this subsection we discuss the neutralino sector 
of \brpv{} at tree-level. We refer to
\citeres{Hempfling:1995wj,Roy:1996bua,Diaz:1997xc,Bisset:1998bt,
Joshipura:1998fn,Hirsch:2000ef,Abada:2001zh,Joshipura:2002fc,
Chun:2002vp,Diaz:2003as,Dedes:2006ni,Liebler:2011tp}
for studies related to neutrino masses in bilinear $R$-parity violation.
If we make use of the basis
\begin{align}
\left( \psi^0 \right)^T &=
\left( {\tilde B}^0, {\tilde W}_3^0, {\tilde H}_d^0, {\tilde H}_u^0,
\nu_1,\nu_2,\nu_3 \right)
\label{eq:basisneut}
\end{align}
the mass matrices of the neutral fermions have the generic form
\begin{align}
\Mass_n^{\text{tree}} = \begin{pmatrix} M_H & \hat{m} \\ \hat{m}^T & 0\end{pmatrix}
\label{eq:neutralmass}
\end{align}
and enter the Lagrangian density as follows
\begin{align}
\label{eq:lagdens}
\Lag\supset -\frac{1}{2}\left((\psi^0)^T\Mass_n^{\text{tree}}\psi^0\right)- \frac{1}{2}\left((\psi^0)^\dagger
\Mass_n^{\text{tree}*}\psi^{0*}\right) \quad.
\end{align}
Therein the sub-matrix $M_H$ is the usual \mssm{} neutralino mass matrix,
whereas the sub-matrix $\hat{m}$ includes the mixing with the left-handed neutrinos
and contains the $R$-parity violating parameters.
In detail the elements are
{\allowdisplaybreaks\begin{align}
M_H = \begin{pmatrix}
M_1 & 0 & -\frac{1}{2}g' v_d & \frac{1}{2}g' v_u & \\
0 & M_2 & \frac{1}{2}g v_d & -\frac{1}{2}g v_u & \\
-\frac{1}{2}g' v_d & \frac{1}{2}g v_d & 0 & -\mu & \\
\frac{1}{2}g' v_u & -\frac{1}{2}g v_u & -\mu & 0 &  \end{pmatrix},\quad
\hat{m}^T = \begin{pmatrix}
-\frac{1}{2} g' v_1 & \frac{1}{2} g v_1 & 0 & \epsilon_1  \\
-\frac{1}{2} g' v_2 & \frac{1}{2} g v_2 & 0 & \epsilon_2  \\
-\frac{1}{2} g' v_3 & \frac{1}{2} g v_3 & 0 & \epsilon_3  \end{pmatrix}
\label{eq:chi0masstree}
\end{align}}\xspaceskip 0pt
with $g$ and $g'$ being the gauge couplings of $SU(2)_L$ and $U(1)_Y$ respectively.
The mass eigenstates $F_i^0$ are related to the gauge eigenstates $\psi_s^0$
by $F_i^0=\mathcal{N}_{is}\psi_s^0$, where the unitary matrix $\Nge$
diagonalizes the full neutralino mass matrix $\Mass_n$ in accordance to
\begin{equation}
\Mass_{n,dia.}=\text{Diag}\left(m_{\tilde{\chi}_1^0},\ldots,m_{\tilde{\chi}_7^0}\right)
=\mathcal{N}^*\mathcal{M}_n^{\text{tree}}\mathcal{N}^\dagger\quad,
\end{equation}
The second part of the Lagrangian density in \eqn{eq:lagdens}
has to be diagonalized by $\mathcal{N}\mathcal{M}_n^{\text{tree}*}\mathcal{N}^T$
in case of \cp{}~violation.
The mass eigenstates in Weyl notation can finally be build up to
$4$-component spinors by
\vspace{-5mm}
\begin{align}
\tilde{\chi}_i^0 = \left( \begin{array}{c}
F_i^0 \\ F_i^{0\dagger}
\end{array} \right)\quad.
\end{align}
The mixing of neutrinos with neutralinos gives rise to one massive
neutrino at tree-level. Its mass yields
\vspace{-5mm}
\begin{flalign}
\label{eq:treelevelmass}
 m_{\nu_3}=\frac{g^2M_1+g'^2M_2}{4\mathrm{det}(M_{H})}|\vec{\Lambda}|^2\quad,
\end{flalign}
with the alignment parameter  $\Lambda_i=\mu v_i + \epsilon_i v_d$.
The atmospheric and the reactor mixing angle of the neutrinos can be expressed
in terms of the alignment parameters $\Lambda_i$ at tree-level by
\begin{flalign}
 \mathrm{tan}^2\theta_{23}=\Bigl|\frac{\Lambda_2}{ \Lambda_3}\Bigr|^2,\qquad
 |U_{e3}|^2\simeq\frac{|\Lambda_1|^2}{|\Lambda_2|^2+|\Lambda_3|^2}\quad.
\end{flalign}
At one-loop level $|U_{e3}|$ can receive considerable corrections.
In the complex case the absolute value of $\Lambda_i$ is given by
$|\Lambda_i|^2 = |\mu v_i + v_d \epsilon_i^R|^2 + |v_d\epsilon_i^I|^2$.
Necessarily the size of $|\epsilon_i^I|\sim |\Lambda_i|/v_d$ is fixed by
the neutrino mass generated at tree-level as shown in \eqn{eq:treelevelmass}
and needs to be smaller than the value of $\epsilon_i^R$ in the pure real case,
where a cancellation between the two terms of $\Lambda_i$ can be arranged.
As a consequence neutrino data
will restrict the size of possible complex phases $\phi_i=\arctan(\epsilon_i^I/\epsilon_i^R)$.
This observation is in accordance to the discussion in \citere{Hirsch:2002tq},
where the cancellation between the terms within $\Lambda_i$ is used to constrain
the complex phases.
Setting the complex phase of $M_2$ to zero allows for a phase~$\phi_\mu$ for
the $\mu$ parameter, which in turn permits larger complex phases for the parameters $\epsilon_i$.
However, $\phi_\mu$ is severely 
constrained by the non-observation of electric dipole-moments and within the allowed range for
$\phi_\mu$ the impact of this phase is small and does not lead to any new qualitative features.

In the following we discuss the effects of \nlo{} corrections
to the neutralino mass matrix, which allow for the explanation of the
solar mass and mixing angle in accordance to \citere{Capozzi:2013csa} as well.
Compared to existing work
\cite{Hempfling:1995wj,Roy:1996bua,Diaz:1997xc,Bisset:1998bt,
Joshipura:1998fn,Hirsch:2000ef,Abada:2001zh,Joshipura:2002fc,
Chun:2002vp,Diaz:2003as,Dedes:2006ni,Liebler:2011tp}
we define $\drbar$ masses at \nlo{} for the neutralino and neutrino sector 
in a slightly different way which is better suited  for the study of \cp{}~violating effects.
The fermionic self-energies can be decomposed as follows
\begin{align}\nonumber
\begin{matrix}\includegraphics[width=0.2\textwidth]{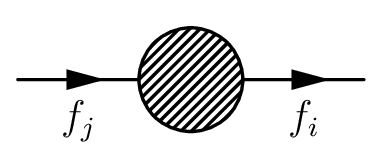}\end{matrix}
\equiv \Gamma_{ij}=\delta_{ij}(\sla{p}-m_{fi})+&\left[\sla{p}\left(P_L\hS_{ij}^{L}\left(p^2\right)
+P_R\hS_{ij}^{R}\left(p^2\right)\right)\right.\\
&\quad\left.+P_L\hS_{ij}^{SL}(p^2)+P_R\hS_{ij}^{SR}(p^2)\right]\quad,
\end{align}
where the hat refers to $\drbar$ renormalized contributions
and $P_{L,R}=\tfrac{1}{2}(1\mp \gamma_5)$ are projection operators.
They enter the Lagrangian
density in the form $-\tfrac{1}{2}\overline{\tilde{\chi}}^0_i\Gamma_{ij}\tilde{\chi}^0_j$
including both terms of \eqn{eq:lagdens}.
In order to calculate $\drbar$ masses for the neutralinos, we have to respect that
$F_i$ and $F_i^*$ are obtained from gauge eigenstates by $\Nge$
and $\Nge^*$ respectively. Taking the different rotations
into account we define the $\drbar$ mass term to be added at \nlo{}
by
\begin{align}
&\Mass_n^{\text{tree}} \rightarrow \Mass_n^{\text{tree}} - \delta \Mass_n \qquad\text{with}\\
\label{eq:drbarnlotermorg}
&\left(\delta \Mass_n\right)_{ij}(p^2)
 =\sum_k \frac{1}{2}(\Mass_n^{\text{tree}})_{ik}(\Nge^\dagger \hS^{R,T}(p^2)\Nge)_{kj}
+\frac{1}{2}(\Nge^T\hS^R(p^2)\Nge^*)_{ik}(\Mass_n^{\text{tree}})_{kj}\\\nonumber
&\qquad\qquad+\frac{1}{2}(\Nge^T\hS^{SL,T}(p^2)\Nge)_{ij} + \frac{1}{2}(\Nge^T\hS^{SL}(p^2)\Nge)_{ij}
\end{align}
As we are dealing with Majorana particles, one finds
\begin{align}
 \Sigma^L_{ij}(p^2)=\Sigma^R_{ji}(p^2),\qquad
 \Sigma^{SL}_{ij}(p^2)=\Sigma^{SL}_{ji}(p^2),\qquad
 \Sigma^{SR}_{ij}(p^2)=\Sigma^{SR}_{ji}(p^2)
\end{align}
such that we are able to rewrite \eqn{eq:drbarnlotermorg}
\begin{align}
\label{eq:drbarnloterm}
 \left(\delta \Mass_n\right)_{ij}(p^2)
 =\sum_k &\frac{1}{2}(\Mass_n^{\text{tree}})_{ik}(\Nge^\dagger \hS^{L}(p^2)\Nge)_{kj}
+\frac{1}{2}(\Nge^T\hS^R(p^2)\Nge^*)_{ik}(\Mass_n^{\text{tree}})_{kj}\\\nonumber
&+(\Nge^T\hS^{SL}(p^2)\Nge)_{ij}\quad.
\end{align}
Therein we perform the following replacement for the practical calculation
\begin{align}
\hS_{ij}(p^2)\rightarrow \frac{1}{2}(\hS_{ij}(m_i^2)+\hS_{ij}(m_j^2))\quad.
\end{align}
\eqn{eq:drbarnloterm} is similar to the formulas in \citeres{Pierce:1996zz,O'Leary:2011yq},
generalized for the left-hand part of \eqn{eq:lagdens}.
Adding the Goldstone tadpoles as done in \citeres{Hirsch:2000ef,Liebler:2010bi,Liebler:2011tp}
allows for the determination of gauge-independent neutralino/neutrino masses at \nlo.
In principle the generalization of the on-shell neutralino and neutrino masses
as shown in \citere{Liebler:2011tp} is straightforward, but will not be presented
in this context. Rather we point out the most important \nlo{} contributions
in the following: For large values of $\tan\beta$ $b$-(s)quark contributions are of quite
importance. However, a major contribution always stems from loops involving a neutral
or charged scalar, whereas loops with gauge bosons are less dominant.

The unitary matrix $\Nge$, which diagonalizes the neutralino mass matrix, contains
the block mixing the neutrino generations, which together with the leptonic block
in the chargino mixing matrices, forms the
Pontecorvo-Maki-Nakagawa-Sakata matrix (\pmns{}) $U^{l\nu}$ matrix~\cite{Maki:1962mu}.
The latter contains at least one \cp{}~violating phase, namely the Dirac phase $\delta$, which enters the
Jarlskog invariant \cite{Jarlskog:1985ht} in the lepton/neutrino sector as follows
\begin{align}
J_{\cp} = \text{Im}(U^{l\nu}_{23}U^{l\nu *}_{13}U^{l\nu}_{12}U^{l\nu *}_{22})
=\frac{1}{8}\cos\theta_{13}\sin(2\theta_{12})\sin(2\theta_{23})\sin(2\theta_{13})\sin\delta\quad.
\label{eq:jarlskog}
\end{align}
We will demonstrate in \sct{sec:neumassesmixing} that in our restricted range of phases $|J_{\cp}|$ can be sizable
and close to the experimental bound.

\subsection{Leptonic neutralino decays}

We turn to the calculation of the decays of the lightest neutralino, which are
dominated by $L$ violating two-body decays for neutralino masses above $m_W$.
We follow \citeres{Liebler:2010bi,Liebler:2011tp}
regarding the evaluation of the decay widths.
Even though the loop contributions, which generate eventually
the lepton asymmetry, are finite, we do perform a complete one-loop analysis
to ensure that the life-time remains long enough such that the neutralino
decays out of equilibrium.
We start with a short discussion of the \lo{} decay
width, for which the relevant part of the Lagrangian density
is given by
{\allowdisplaybreaks\begin{align}
\begin{split}
\mathcal{L}\supset &
\overline{\tilde{\chi}_l^0}\gamma^\mu
\left(P_LO_{Llj}^Z+P_RO_{Rlj}^Z\right)
\tilde{\chi}_j^0 Z_\mu +
\left(\overline{\tilde{\chi}_l^-}\gamma^\mu
\left(P_LO_{Llj}^W+P_RO_{Rlj}^W\right)
\tilde{\chi}_j^0 W^-_\mu +
 \text{h.c.}\right) \\ 
 &  + \overline{\tilde{\chi}_l^0} (P_LO_{Llj}^{h^0}+P_RO_{Rlj}^{h^0}) \tilde{\chi}_j^0 h^0\quad.
\label{eq:treelevellag}
\end{split}
\end{align}}\xspaceskip 0pt
It includes the coupling to the charged leptons $l^\pm$, which are
part of the charginos $\tilde{\chi}_l^\pm$. 
The explicit form of the couplings can be taken from \citere{Liebler:2011tp}.
The widths for the channels involving a final state gauge boson can be written as follows
\begin{equation}
\Gamma^{0} = \frac{\sqrt{\kappa(m_i^2,m_o^2,m_V^2)}}{16\pi m_i^3}
\left[\left(|O_{L}^V|^2+|O_{R}^V|^2 \right) f(m_i^2,m_o^2,m_V^2)
-6 \mathrm{Re}(O_{L}^VO_{R}^{V*})m_im_o\right]
\label{eq:leadingorderwidth}
\end{equation}
with $V\in \lbrace W,Z\rbrace$, the masses of the mother (daughter) particle $m_i$ ($m_o$)
and the functions
\begin{align}
f(x,y,z) = \frac{x+y}{2}- z+\frac{(x-y)^2}{2 z}, \quad
\kappa(x,y,z)=x^2+y^2+z^2-2xy-2xz-2yz\quad.
\label{eq:kaellenfunction}
\end{align}
In case of the channel with a final state Higgs boson the partial width is given by
\begin{equation}
\Gamma^{0} = \frac{\sqrt{\kappa(m_i^2,m_o^2,m_h^2)}}{16\pi m_i^3}
\left[\frac{|O_{L}^{h^0}|^2+|O_{R}^{h^0}|^2}{2}
(m_i^2+m_o^2-m_h^2) + 
2\mathrm{Re}(O_{L}^{h^0}O_{R}^{h^0*})m_im_o\right] \,.
\end{equation}

The \lo{} decay widths with leptons/neutrinos and antileptons/antineutrinos 
in the final state
are identical. In order to observe \cp{}~violating effects with respect to
the different final states we proceed as in \citeres{Liebler:2010bi,Liebler:2011tp}
and calculate \nlo{} contributions, which are all implemented in \prog{}.
The \nlo{} decay widths can be written in the form
\begin{align}
 \Gamma^{1}=\Gamma^0 + \frac{\sqrt{\kappa(m_i^2,m_o^2,m_V^2)}}{16\pi m_i^3}
 \frac{1}{2}\sum_{\text{pol}} 2\rm{Re}\left(M_1M_0^\dagger\right)
\end{align}
with the tree-level amplitude $M_0$ and the \nlo{} amplitude $M_1$. The latter
includes the \nlo{} vertex corrections as well as the wavefunction corrections
of in- and outgoing particles as discussed in \citeres{Liebler:2010bi,Liebler:2011tp}.
For the decay $\chiN\rightarrow l^\pm W^\mp$ real corrections by photon emission
are added accordingly. For $\chiN\rightarrow \nu(\overline{\nu})Z$
we distinguish neutrinos and antineutrinos as follows:
We assign a lepton number $+1 (-1)$ to the left-handed (right-handed) part of the
neutrino Dirac spinor, which due to the smallness of neutrino masses
and the energies considered here is an extremely good approximation.
For \lsp{} masses above the lightest Higgs mass the
decay channel $\chiN\rightarrow \nu(\bar{\nu})h$ is relevant as well.
We implemented the \nlo{} virtual contributions for the latter decays to \prog{} in order to
estimate the \cp{}~asymmetry with respect to the different final
states and find a similar asymmetry as in the decays involving heavy gauge boson
final states.

\begin{figure}[t]
\begin{center}
 \begin{tabular}{ccc}
\includegraphics[width=0.28\textwidth]{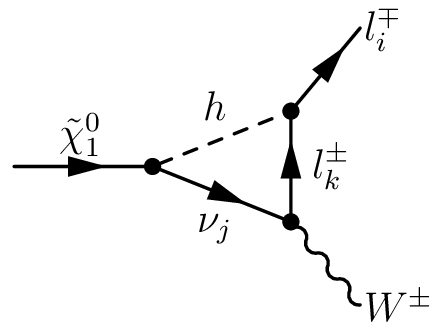} &
\includegraphics[width=0.28\textwidth]{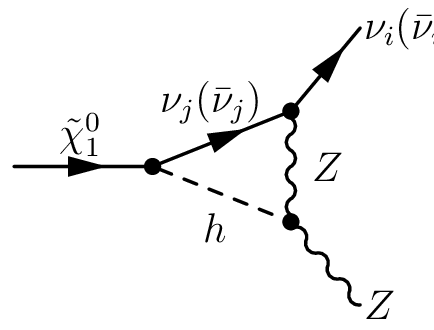} &
\includegraphics[width=0.28\textwidth]{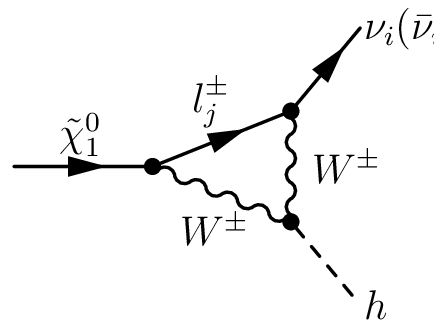}\\[-0.3cm]
 (a) & (b) & (c)
\end{tabular}
\end{center}
\vspace{-5mm}
\caption{Dominant \nlo{} virtual contribution generating a \cp{}~asymmetry
between the different final states for
(a) $\chiN\rightarrow l^\pm W^\mp$;
(b) $\chiN\rightarrow \nu Z$;
(c) $\chiN\rightarrow \nu h$.}
\label{fig:nlocont}
\end{figure}

In \fig{fig:nlocont} we show the dominant \nlo{} virtual contributions, which
generate the \cp{}~asymmetry between the final states $l^-W^+$ and $l^+W^-$
in accordance to \citere{Hirsch:2002tq}, $\nu Z$ and $\bar{\nu}Z$ as well as
$\nu h$ and $\bar{\nu} h$. In case of a light stau, the corresponding light
stau loop contribution to final states containing 
a (anti-)neutrino could be as important as the ones shown.

\subsection{Number density evolution via Boltzmann equations}

Within this section we present the evolution of number densities in the universe
at temperatures which correspond to energies around the electroweak scale. For a quantitative
discussion we make use of Boltzmann equations, in which we take into account the decays
as well as the inverse decays of the lightest neutralino.
Moreover we add $R$-parity conserving annihilation processes of the \lsp{}, which
are known to have an impact on the final particle densities  \cite{Hambye:2001eu}.
Sphaleron transitions between baryon and lepton asymmetries are discussed
in \sct{sec:theobaryo}.
However, we can neglect $R$-parity violating scattering processes 
changing the lepton number by one or two units, since those processes
involve an intermediate neutrino or neutralino and are thus
either suppressed by the small neutrino mass or a product of $R$-parity violating
couplings. In addition \cp{}~violating scatterings affect final particle densities
only slightly, if the neutralino density stays close to its equilibrium density
as pointed out in \citere{Buchmuller:2004nz}.
Thus, the Boltzmann equations take the generic form 
\begin{align}
\begin{split}
xH(x)\frac{dN_{\tilde{\chi}_1^0}}{dx}= -\underset{i,j}{\sum} \Bigl [ &\frac{\mathrm{K}_1(x)}{\mathrm{K}_2(x)}
 \Bigl (N_{\tilde{\chi}_1^0}\Gamma(\tilde{\chi}_1^0\, {\to}\,ij)- \frac{N_iN_j}{N_{i}^{eq}N_{j}^{eq}}
N_{\tilde{\chi}_1^0}^{eq}\Gamma(ij\,{\to}\,\tilde{\chi}_1^0) \Bigr) \\
 & + \hat{\sigma}(\tilde{\chi}_1^0\tilde{\chi}_1^0\, {\to}\,i j) 
\Bigl (N^2_{\tilde{\chi}_1^0}-\frac{N_iN_j}{N_{i}^{eq}N_{j}^{eq}}N^{eq,2}_{\tilde{\chi}_1^0} \Bigr)\Bigr ]\quad,
\end{split}
\label{eq:neutralinodensevol2}
\end{align}
where $i,j$ denote \sm{} particles.
$\mathrm{K}_1(x)$ and $\mathrm{K}_2(x)$ are modified Bessel functions,
the parameter $x=m_{\tilde{\chi}_1^0}/\mathrm{T}$ denotes the inverse of the temperature
and $\Gamma$ is the usual decay width
in the rest frame of the decaying particle.\footnote{We are using units
where the Boltzmann constant $k_B$ is set to 1.} Moreover we define the
density~$N_i:=N_i(x)=n_i(x)/s(x)$ per co-moving volume element by the
ratio of the particle density $n_i(x)$ to the entropy $s(x)$.
The quantity $\hat{\sigma}$ contains the thermally averaged annihilation
cross section $\langle\sigma_{ij} v \rangle$ of the \lsp{}  \cite{Belanger:2004yn}
\begin{align}
 \hat{\sigma}(\chiN \chiN \rightarrow ij)= x H(x)\frac{m_{\chiN}}{x^2}\sqrt{\frac{\pi g_*}{45}} M_p \langle\sigma_{ij} v \rangle
 \quad\text{with}\quad H:=H(x) = \sqrt{\frac{4\pi^3g_*}{45}}\frac{m_{\chiN}^2}{M_p}\frac{1}{x^2}\quad.
\end{align}
The latter formulas include the Planck mass $M_p$ and the effective degrees of freedom~$g_*$,
which are taken as a function of $x$ from the tabulated values in {\tt micrOMEGAs} \cite{Belanger:2013oya}.
The thermally averaged cross section $\langle\sigma_{ij} v \rangle$ can be calculated with
the help of {\tt micrOMEGAs} as the $R$-parity violating parameters are too small to impact
on the \mssm{} annihilation cross sections.

Since on cosmological timescales the massive gauge bosons and the Higgs boson
decay instantaneously,
we directly elaborate the Boltzmann equations with the decay products,
which assumes the validity of the narrow-width approximation.
In turn \eqn{eq:neutralinodensevol2} can be written in the form
\begin{flalign}
 \begin{split}
xH\frac{dN_{\tilde{\chi}_1^0}}{dx} = & -\frac{\mathrm{K}_1(x)}{\mathrm{K}_2(x)}
\underset{i,q,\bar{q}}{\sum} 
   \Bigl[ N_{\tilde{\chi}_1^0}\Gamma(\tilde{\chi}_1^0\, {\to}\,\bar\nu_{i}Z)\mathrm{Br}(Z\, \to \, q\bar{q}) \\
& \qquad - \frac{N_{\bar\nu_i}}{N_\nu^{eq}} \frac{N_{q} N_{\bar{q}}}{N^{eq}_q N^{eq}_q}
N_{\tilde{\chi}_1^0}^{eq}\Gamma(\bar\nu_{i} Z\,{\to}\,\tilde{\chi}_1^0) 
 \mathrm{Br}(Z\, \to \, q\bar{q})+ \dots \Bigr] \\
& - \underset{q,\bar{q}}{\sum} \Bigl[\hat{\sigma}(\tilde{\chi}_1^0\tilde{\chi}_1^0  \, \to q \bar{q}) 
 (N^2_{\tilde{\chi}_1^0}-\frac{N_{q} N_{\bar{q}}}{N_q^{eq}N_{q}^{eq}} {N_{\tilde{\chi}_1^0}^{eq}}^2) \Bigr]+\dots\quad,
 \end{split}
\end{flalign}
where we have just presented the decay of the heavy gauge boson to a quark pair in combination with
the \lsp{} annihilation process to this specific final state. The complete set of
formulae is given in Appendix~\ref{sec:formulas}.
To shorten our notation we sum up the generations of $u$- and $d$-type quarks
and denote them $q_1$ and $q_2$ in our study. The Boltzmann
equations for the number densities of the (anti-)leptons,
(anti-)neutrinos and (anti-)quarks are obtained similarly
and presented in Appendix~\ref{sec:formulas} as well.

\subsection{Baryogenesis via Leptogenesis}
\label{sec:theobaryo}

The lepton asymmetry can be transformed into a baryon asymmetry and vice versa via
sphaleron transitions \cite{Khlebnikov:1988sr,Harvey:1990qw,Hambye:2001eu}.
However, the sphaleron rate is dramatically suppressed
for temperatures below the electroweak scale, thus after the electroweak
phase transition. We follow \citeres{D'Onofrio:2012jk,D'Onofrio:2014kta}, which discuss the sphaleron
rate in the light of the recent Higgs discovery with $m_H\sim 125$\,GeV leading
to a critical temperature of $T_c=159\pm 1$\,GeV. Due to the fast
drop of the sphaleron rate for temperatures below $T_c$, we can safely assume
that the baryon asymmetry decouples from the lepton asymmetry at this temperature.
Effects resulting from the transition region down to temperatures of $m_{\chiN}\sim 100$\,GeV
are tiny with respect to our qualitative discussion.
Nonetheless we implemented formulas (1.10) and (1.11) of \citere{D'Onofrio:2012jk}
and split them accordingly to particles and antiparticles.
For this purpose
we define the lepton asymmetry $\delta_N$ as sum over neutrino and lepton flavors by
\begin{align}
\delta_N = \sum_{i=1,2,3}  N_{l^-_i}- N_{l^+_i}+ N_{\nu_i}-N_{\bar \nu_i}
\label{eq:leptonasymmetry}
\end{align}
and accordingly the baryon asymmetry in the form
\begin{align}
\delta_B = N_{q_1} - N_{\bar q_1} + N_{q_2} - N_{\bar q_2}\quad.
\end{align}
Formulas (1.10) and (1.11) of \citere{D'Onofrio:2012jk} then read e.g.
\begin{align}
 xH\frac{d(N_{q_1}-N_{\bar q_1})}{dx}=\frac{\gamma(x)}{2}[\delta_B+\eta(x)\delta_N],\quad
 xH\frac{d(N_{l_i^-}-N_{l_i^+})}{dx} =\frac{\gamma(x)}{6}[\delta_B+\eta(x)\delta_N].
\end{align}
The function $\gamma(x)$ incorporates the strength of the sphaleron transitions
and thus drops rapidly to zero for $T<T_c$, i.e. $x>m_{\chiN}/T_c$.
The function $\eta(x)$ determines the ratio of $\delta_B$ and $\delta_N$ for $x<T_c/m_{\chiN}$.
We use $\eta(x)= 0.5$, which is a reasonable approximation for our study.
The corresponding results of this procedure are presented in \sct{sec:numbaryo}.

\section{Numerical results}
\label{sec:results}

In this section we show numerical results obtained with the previously
discussed formulas and tools. We first stick to the \cp{}~conserving case of
\brpv{}, before discussing the effects of \cp{}~violation on lepton asymmetries
and baryon asymmetries in the early universe.
Our discussion is based on the following low-energy \susy{} points: 
The soft-breaking masses are set diagonal to $M_L=M_E=1$~TeV
and $M_Q=M_U=M_D=1.5$~TeV (generation independent)
and the gaugino masses are fixed to $M_1=100$~GeV, $M_2=400$~GeV, $M_3=1.5$~TeV, which also ensures
compatibility with the latest ATLAS results \cite{Aad:2014pda}. We choose the soft-breaking couplings
to be  $A_b=-1$~TeV and $A_\tau=-500$~GeV. 
We finally define two points with small and large value of $\tan\beta$, namely:
\begin{align}
\text{Scenario $P_1$:} \,\, \tan\beta=5\,,\,A_t=3\,\text{TeV and Scenario $P_2$:} 
\,\, \tan\beta=35\,,\,A_t=2.5\,\text{TeV}
\end{align}
Similar to the lepton sector the soft-breaking masses of the squark sector
are set diagonal. In both cases we set $\mu=1$~TeV and $m_A=370$~GeV resulting in
a lightest \sm-like Higgs\,$h$ with mass $m_h$ close to 125~GeV and a lightest neutralino 
with mass close to $105$~GeV\footnote{For completeness we note, that
we have calculated the Higgs masses in the $R$-parity conserving limit
including 2-loop effects using {\tt SPheno} \cite{Porod:2003um,Porod:2011nf} 
as the complete formulae for the 
Higgs masses including $R$-parity violation are not known. However, this
is an excellent approximation as the 
corresponding couplings are much smaller than the $R$-parity conserving ones.}.

\subsection{Neutrino masses and mixing angles}
\label{sec:neumassesmixing}

Neutrino data provides a constraint on the possible phases of the parameters~$\epsilon_i$,
which can be easily understood by
$|\Lambda_i|^2 = |\mu v_i + v_d \epsilon_i^R|^2 + |v_d\epsilon_i^I|^2$.
As discussed in \sct{sec:neutrinomassestree} in the real \brpv{} a cancellation in the sum
$\mu v_i + v_d \epsilon_i$ allows to explain the atmospheric neutrino mass scale
at tree-level on the one hand side together with an explanation of the solar mass
scale at the loop level thanks to sufficiently large $\epsilon_i$ on the other hand
\cite{Hirsch:2000ef,Diaz:2003as}.
As a consequence purely imaginary $\epsilon_i$ for all generations at the same time
are impossible. Moreover
as $v_d$ decreases with increasing $\tan\beta$ we expect that larger phases are possible
for larger values of $\tan\beta$.

\begin{figure}[htp]
\begin{center}
\begin{tabular}{cc}
\includegraphics[width=0.48\textwidth]{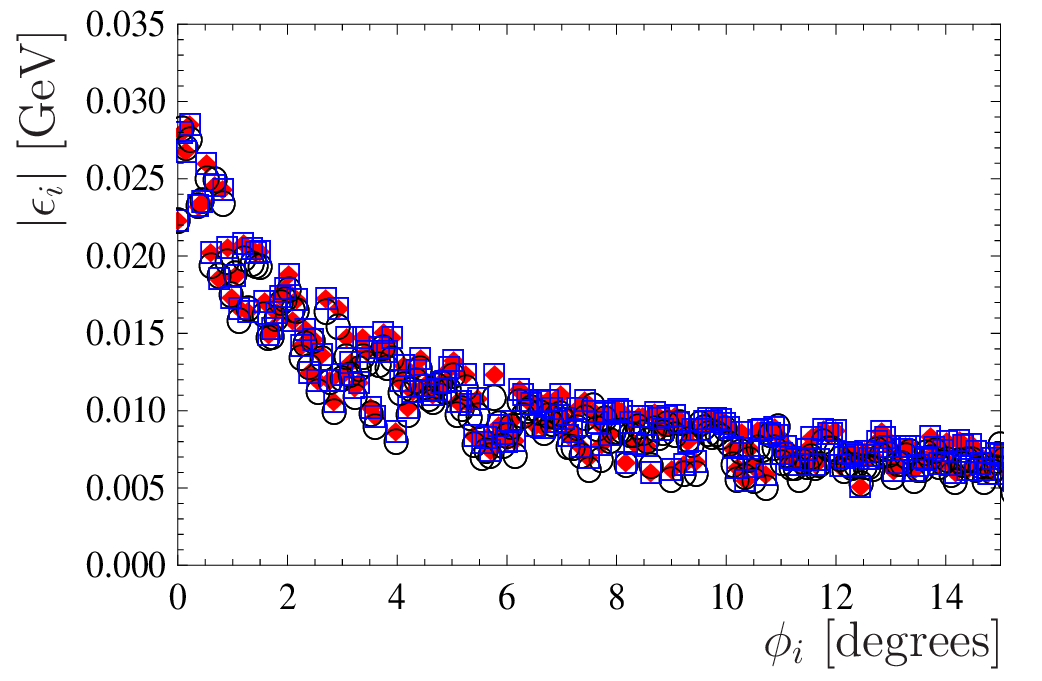} &
\includegraphics[width=0.48\textwidth]{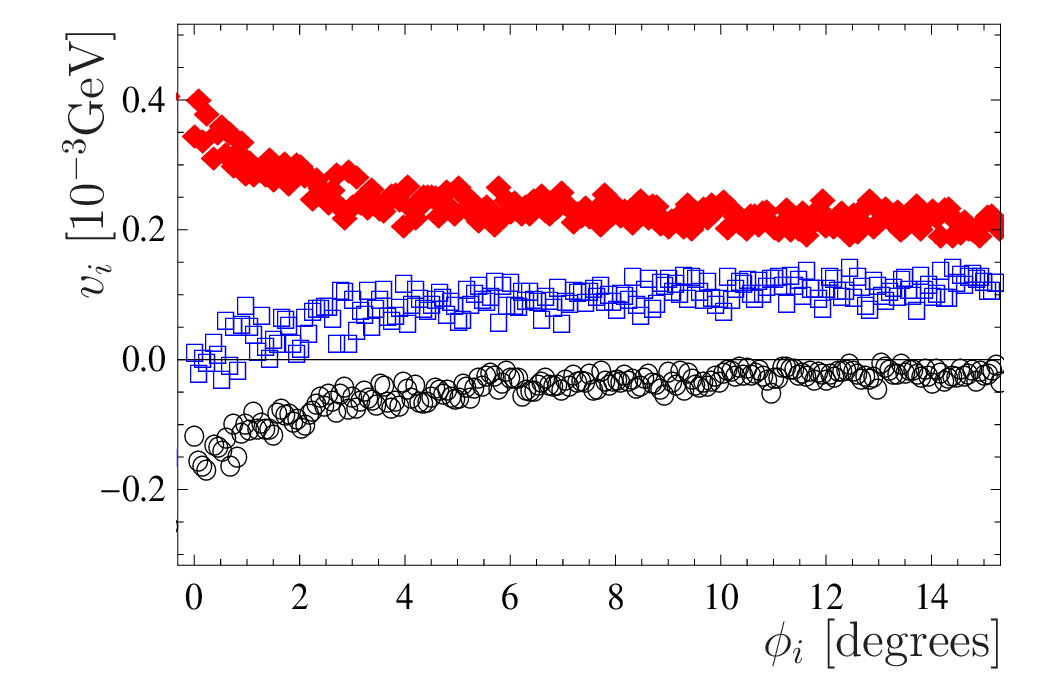}  \\[-0.3cm]
 (a) & (b)
\end{tabular}
\end{center}
\vspace{-5mm}
\caption{Values of $R$-parity violating parameters as a function of 
$\phi:=\phi_1=\phi_2=\phi_3$  in degrees for scenario $P_2$, where in 
(a) we give $|\epsilon_i|$ in GeV with $i=1$ (black, circle),
$2$ (blue, square) and $3$ (red, diamond) and in
(b) $v_i$ in $10^{-3}$~GeV with the same coding.}
\label{fig:epsvevL-epsphase}
\begin{center}
\begin{tabular}{cc}
\includegraphics[width=0.48\textwidth]{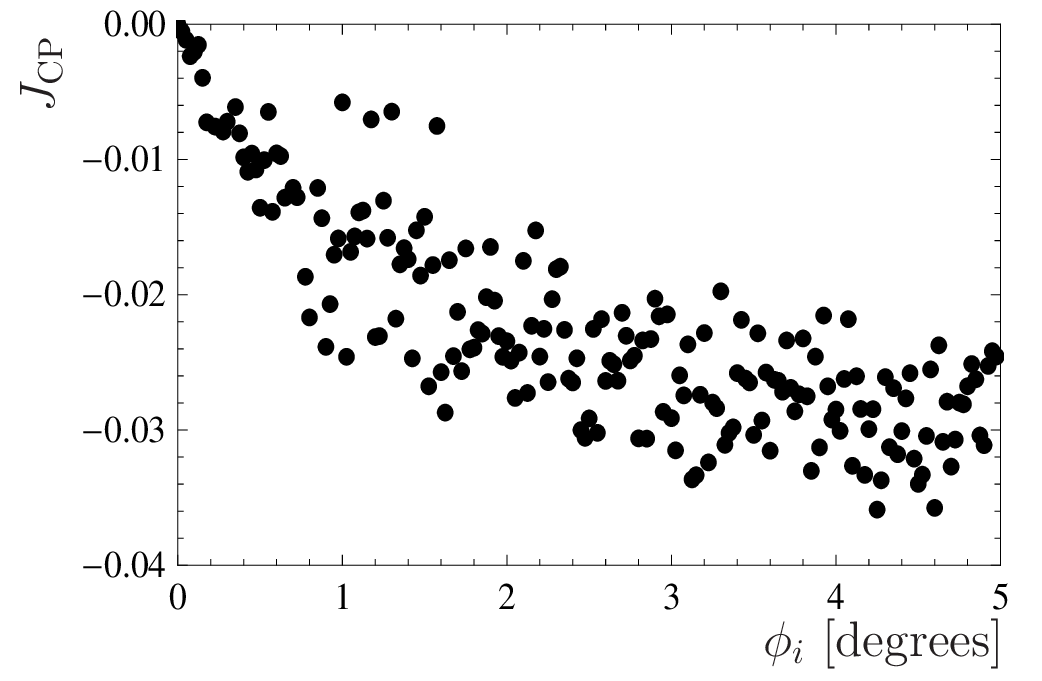} &
\includegraphics[width=0.48\textwidth]{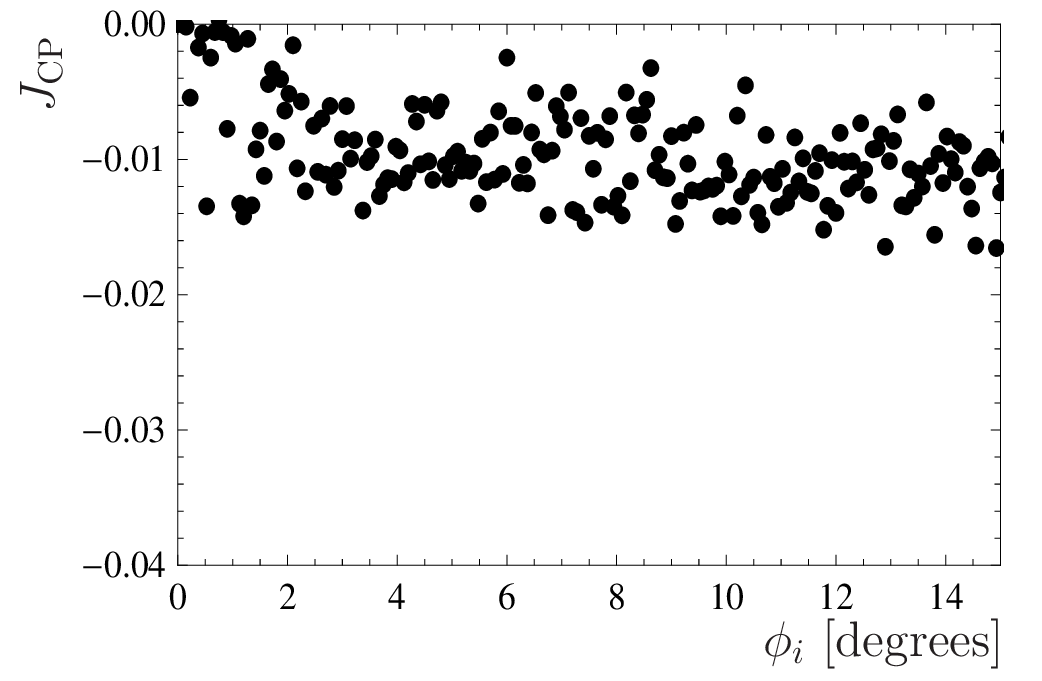}  \\[-0.3cm]
 (a) & (b)
\end{tabular}
\end{center}
\vspace{-5mm}
\caption{$J_{\cp}$ as a function  $\phi:=\phi_1=\phi_2=\phi_3$ in degrees
for (a) $P_1$ and (b) $P_2$.}
\label{fig:JCP-epsphase}
\end{figure}

As an example we give in \fig{fig:epsvevL-epsphase} the adjustment of $v_i$ and the
absolute values of $\epsilon_i$ as a function of the
phases $\phi_i=\arctan\left(\epsilon_i^I/\epsilon_i^R\right)$ for the case  $P_2$.
Here and in the following we will take the phases of all three $\epsilon_i$ to be equal to 
maximize the effects. The width of the bands
reflects the experimental uncertainty of the neutrino data and the upper bound
of the complex phases is given
by the requirement to obtain correctly both neutrino mass scales at the same time. 
In principle one could get a somewhat larger range be adjusting the soft parameters in
the sbottom and in the stau sector  \cite{Hirsch:2000ef,Diaz:2003as}.
However, as no new features show up for larger values of the complex phases
we do not pursue this road. This can also be seen by checking the Jarlskog invariant $J_{\cp}$ 
of the \pmns{}-matrix defined in \eqn{eq:jarlskog} which we show
in \fig{fig:JCP-epsphase}. Taking the current neutrino data
leads to an upper bound $|J_{\cp}|\leq 0.040$
assuming a maximal Dirac phase. Note, that we reach this bound in case of scenario $P_1$.

\subsection{Decay of the \lsp{}}
\label{sec:neudecays}

The smallness of the neutrino masses and in turn the smallness of the \brpv{} parameters
imply a small decay width of the neutralino, such
that it decays out of equilibrium in the early universe.
Additionally the decays come with displaced vertices in collider experiments
\cite{deCampos:2007bn,DeCampos:2010yu,deCampos:2012pf} 
allowing an experimental verification of this feature at the \lhc{}.
Taking the real values for the $R$-parity breaking parameters as provided in
\tab{tab:rp_parameters} we find for scenario~$P_1$  at \nlo{}
\begin{align}
&\Gamma\left(\tilde{\chi}_1^0\rightarrow e^\pm W^\mp\right)=3.36\cdot 10^{-16}~\mathrm{GeV},
&&\Gamma\left(\tilde{\chi}_1^0\rightarrow \mu^\pm W^\mp\right)=1.31\cdot 10^{-14}~\mathrm{GeV}\\
&\Gamma\left(\tilde{\chi}_1^0\rightarrow \tau^\pm W^\mp\right)=1.44\cdot 10^{-14}~\mathrm{GeV},
&&\Gamma\left(\tilde{\chi}_1^0\rightarrow \nu_3(\overline{\nu}_3) Z\right)=3.27\cdot 10^{-15}~\mathrm{GeV}
\end{align}
and  for scenario $P_2$ accordingly
\begin{align}
&\Gamma\left(\tilde{\chi}_1^0\rightarrow e^\pm W^\mp\right)=3.07\cdot 10^{-16}~\mathrm{GeV},
&&\Gamma\left(\tilde{\chi}_1^0\rightarrow \mu^\pm W^\mp\right)=7.71\cdot 10^{-15}~\mathrm{GeV}\\
&\Gamma\left(\tilde{\chi}_1^0\rightarrow \tau^\pm W^\mp\right)=1.04\cdot 10^{-14}~\mathrm{GeV},
&&\Gamma\left(\tilde{\chi}_1^0\rightarrow \nu_3(\overline{\nu}_3) Z\right)=2.35\cdot 10^{-15}~\mathrm{GeV}\quad.
\end{align}

\begin{table}[htp]
\begin{center}
\begin{tabular}{|l||r|r|r|}\hline 
scenario & \multicolumn{1}{c|}{$P_1$} & \multicolumn{1}{c|}{$P_2$}
   & \multicolumn{1}{c|}{$P_1'$} \\ \hline\hline
$\epsilon_1$\,[GeV] & $3.12\cdot 10^{-2}$  & $2.21\cdot 10^{-2}$  & $2.62\cdot 10^{-2}$\\
$\epsilon_2$\,[GeV] & $3.13\cdot 10^{-2}$  & $2.22\cdot 10^{-2}$  & $2.62 \cdot 10^{-2}$\\
$\epsilon_3$\,[GeV] & $-3.13\cdot 10^{-2}$  & $-2.22\cdot 10^{-2}$ & $-2.62\cdot 10^{-2}$\\\hline
$v_1$\,[GeV]        & $-1.45\cdot 10^{-3}$ & $-1.20\cdot 10^{-4}$ & $-1.23\cdot 10^{-3}$\\
$v_2$\,[GeV]        & $-1.27\cdot 10^{-3}$  & $7.70\cdot 10^{-6}$  & $-1.05\cdot 10^{-3}$\\
$v_3$\,[GeV]        & $1.71\cdot 10^{-3}$  & $3.40\cdot 10^{-4}$  & $1.43\cdot 10^{-3}$\\ \hline
\end{tabular}
\end{center}
\caption{Standard choice of real $R$-parity violating parameters for the three scenarios
such that neutrino data \cite{Capozzi:2013csa} are correctly explained.}
\label{tab:rp_parameters}
\end{table}

The  final states $e^\pm W^\mp$ have considerably smaller decay widths than the others,
the reason being the generation of the atmospheric scale of neutrino mixing at tree-level.
The decays in the two lightest neutrino flavors are both vanishing at tree-level \cite{Porod:2000hv}, but
also at loop-level. Due to $\Gamma < H(T=m_{\chiN})\approx 2\cdot 10^{-14}$\,GeV the decay
of the lightest neutralino at the temperature $T=m_{\chiN}$ occurs out of equilibrium.
Further details with respect to the out-of-equilibrium decay can be found in the
subsequent section. We emphasize that an out-of-equilibrium decay only occurs
for a light neutralino $m_{\tilde\chi^0_1}\lesssim 100$\,GeV. 
For heavier neutralinos $\gtrsim 150$\,GeV
the decay widths start to exceed the Hubble parameter. Only a detailed
description with Boltzmann equations can reveal the impact on lepton and baryon
asymmetries.

\begin{figure}[htp]
\begin{center}
\begin{tabular}{cc}
\includegraphics[width=0.48\textwidth]{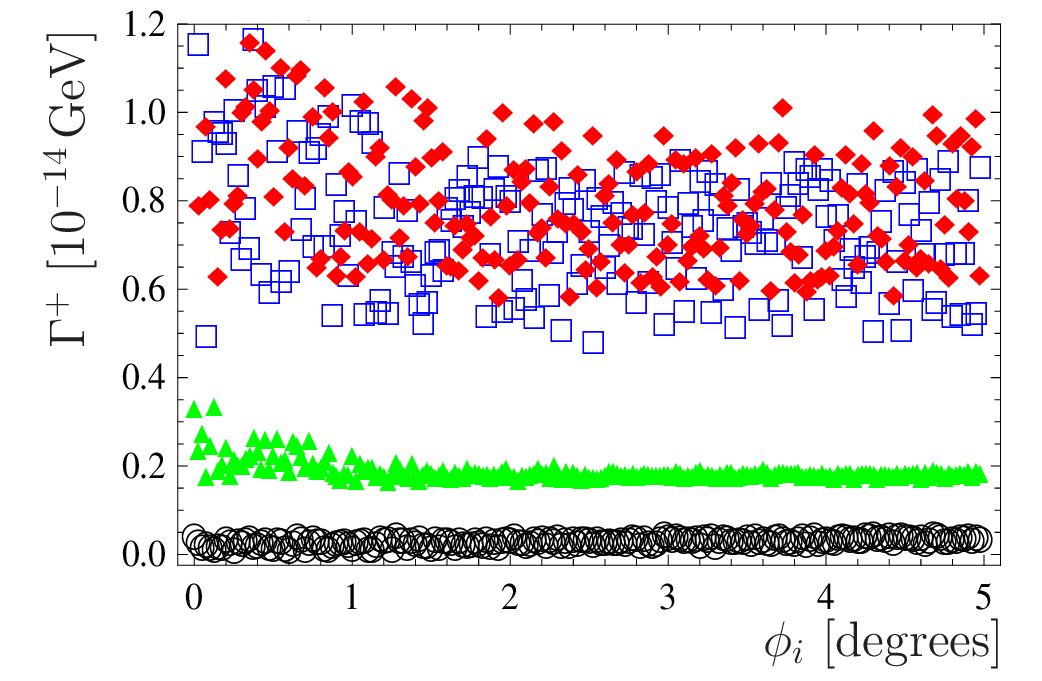} &
\includegraphics[width=0.48\textwidth]{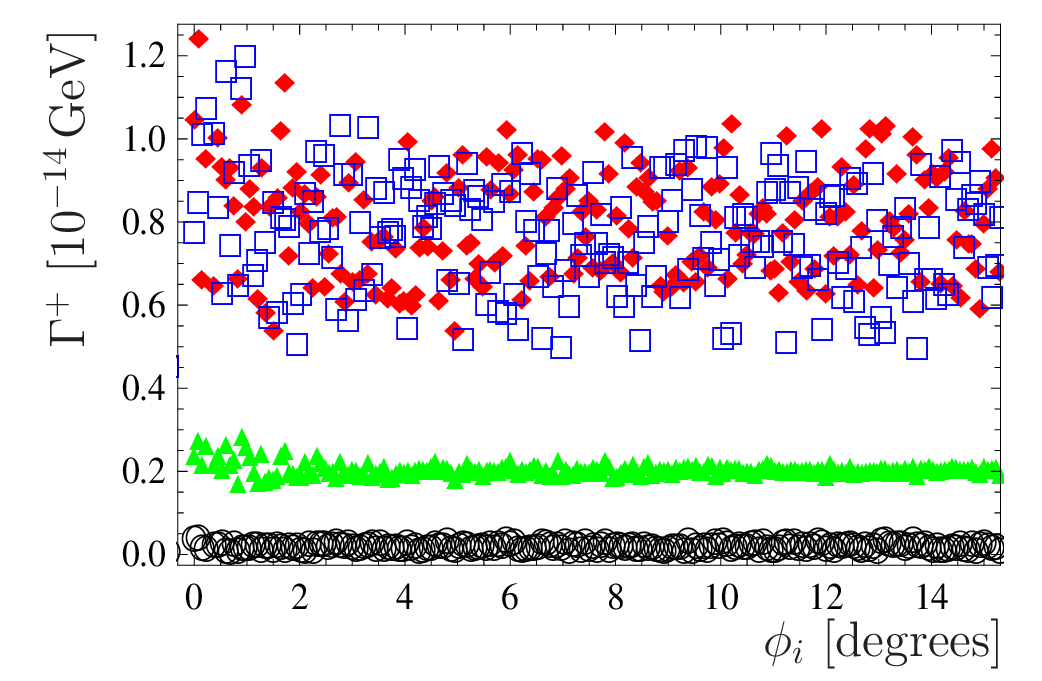}  \\[-0.3cm]
 (a) & (b)
\end{tabular}
\end{center}
\vspace{-5mm}
\caption{Decay widths $\Gamma^+$ for the final states $e^+W^-$ (black, circle), $\mu^+W^-$ (blue, square),
$\tau^+W^-$ (red, diamond), $\bar{\nu}_3Z$ (green, triangle)
as a function of $\phi:=\phi_1=\phi_2=\phi_3$ in degrees for
(a) $P_1$ and (b) $P_2$.}
\label{fig:complexwidths}
\vspace{-5mm}
\begin{center}
\begin{tabular}{cc}
\includegraphics[width=0.48\textwidth]{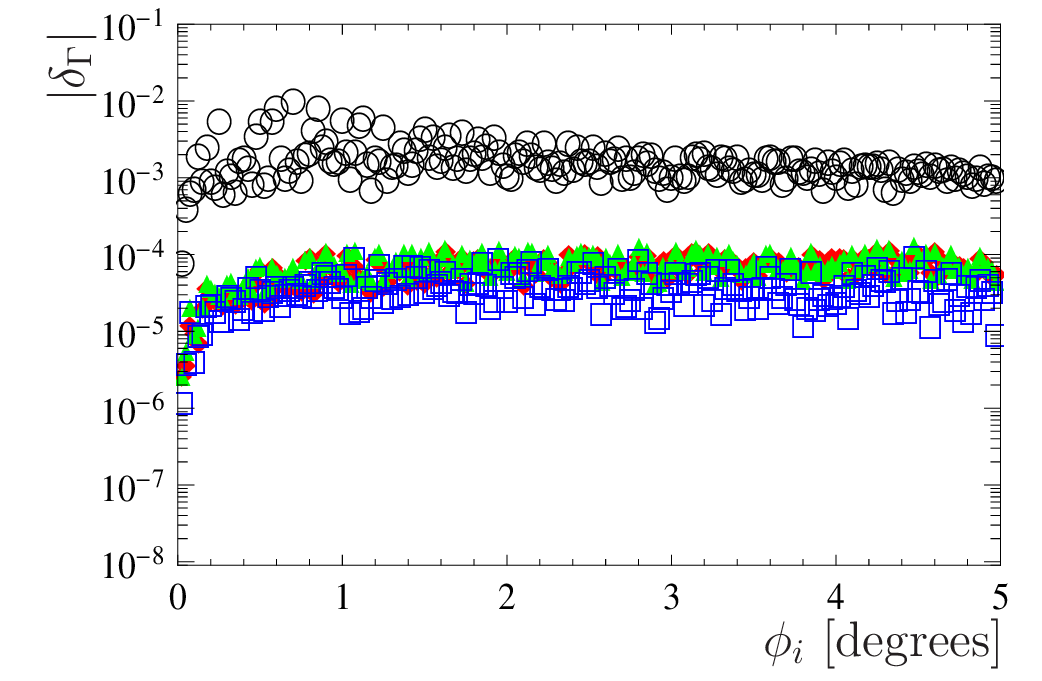} &
\includegraphics[width=0.48\textwidth]{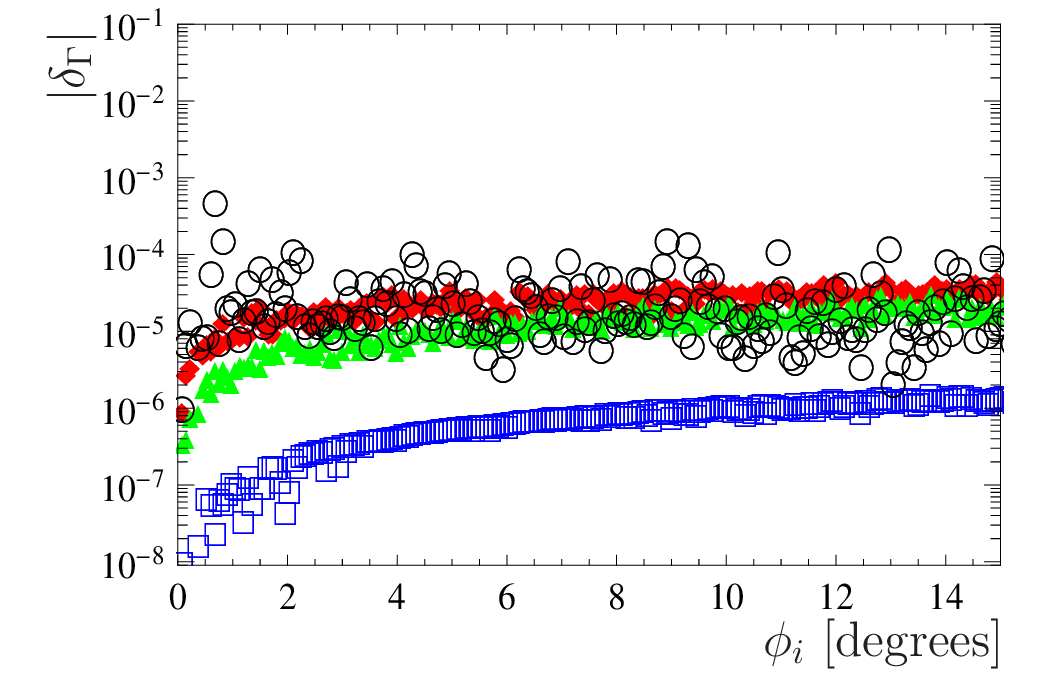}  \\[-0.3cm]
 (a) & (b)
\end{tabular}
\end{center}
\vspace{-5mm}
\caption{\cp{}~asymmetry $|\delta_\Gamma|$ for the final states
$e^\pm W^\mp$ (black, circle), $\mu^\pm W^\mp$ (blue, square),
$\tau^\pm W^\mp$ (red, diamond), $\nu_3(\bar{\nu}_3)Z$ (green, triangle)
defined in \eqn{eq:widthasymmetry} as a function of $\phi:=\phi_1=\phi_2=\phi_3$ in degrees for
(a) $P_1$ and (b) $P_2$.}
\label{fig:complexasymmetry}
\end{figure}

Before presenting our results for the particle densities in the early universe, let us
briefly mention the effects of the complex phases $\phi_i$ on the partial decay widths and
the individual \cp{}~asymmetries, the latter being defined as
\begin{align}
 \delta_\Gamma = \frac{\Gamma^+-\Gamma^-}{\Gamma^++\Gamma^-}
 \label{eq:widthasymmetry}
\end{align}
with $\Gamma^\pm = \Gamma(\tilde{\chi}_1^0\rightarrow l^\pm W^\mp)$ or
$\Gamma^\pm = \Gamma(\tilde{\chi}_1^0\rightarrow \nu(\overline{\nu}) Z)$.
For both scenarios $P_1$ and $P_2$ \fig{fig:complexwidths} shows the decay widths $\Gamma^+$
of the \lsp{} in GeV for the various decay channels.
The \cp{}~asymmetry~$\delta_\Gamma$ is presented in \fig{fig:complexasymmetry} for both scenarios:
The \cp{}~violation in the final state involving electrons is effectively of the same size
as in the other decay channels, if one takes into account the smallness of the decay widths
in this particular final state.
Changing the sign $\sin\phi_i\rightarrow -\sin\phi_i$ for all $i\in\lbrace 1,2,3\rbrace$
yields $\delta_\Gamma\rightarrow -\delta_\Gamma$.

\begin{figure}[ht]
\begin{center}
\includegraphics[width=0.48\textwidth]{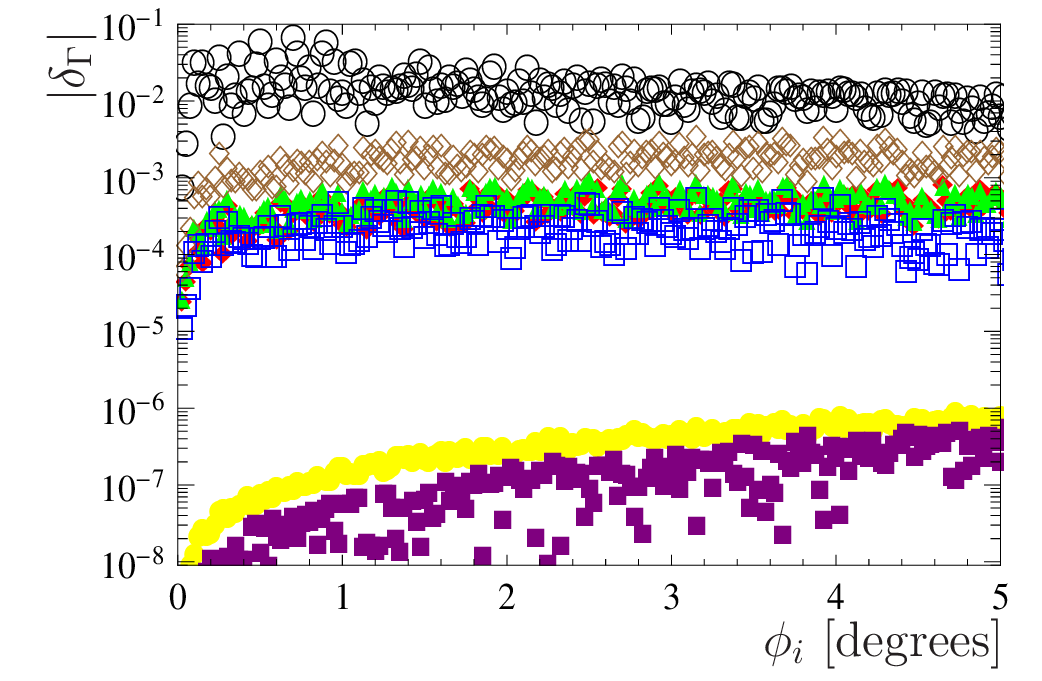}
\vspace{-3mm}
\caption{\cp{}~asymmetry $|\delta_\Gamma|$ for the final states $e^\pm W^\mp$ (black, empty circle),
$\mu^\pm W^\mp$ (blue, empty square),
$\tau^\pm W^\mp$ (red, filled diamond),
$\nu_3(\bar{\nu}_3)Z$ (green, filled triangle),
$\nu_1(\bar{\nu}_1)h$ (yellow, filled circle),
$\nu_2(\bar{\nu}_2)h$ (purple, filled square)
and $\nu_3(\bar{\nu}_3)h$ (brown, empty triangle)
defined in \eqn{eq:widthasymmetry} as a function of $\phi:=\phi_1=\phi_2=\phi_3$ in degrees for $P'_1$.}
\label{fig:complexasymmetry_2}
\end{center}
\end{figure}

In order to estimate the size of the asymmetry in decays $\tilde{\chi}_1^0\rightarrow \nu(\bar\nu)h$
we add a scenario~$P'_1$, where the gaugino mass $M_1$ is shifted from $100$\,GeV to $150$\,GeV, such that the
decay channel into the light Higgs $h$ opens. A possible set of $R$-parity violating parameters
fulfilling neutrino data is added to \tab{tab:rp_parameters} and \fig{fig:complexasymmetry_2}
presents the corresponding \cp{}~asymmetries. The \cp{}~asymmetry in the final state $\nu_3(\bar{\nu}_3)h$ is of a similar
size as in case of the final states with gauge boson, whereas the ones in the first two
neutrino mass generations are negligible. The $\nu_i(\bar \nu_i)h$ final states have
branching fractions comparable to $\mu^\pm/\tau^\pm W^\mp$ final states and thus their inclusion
for $m_{\chiN}>m_h$ is advisable.

\subsection{Lepton asymmetries in the \cp{}~conserving \brpv{}}

\begin{figure}[ht]
\begin{center}
\begin{tabular}{cc}
\includegraphics[width=0.48\textwidth]{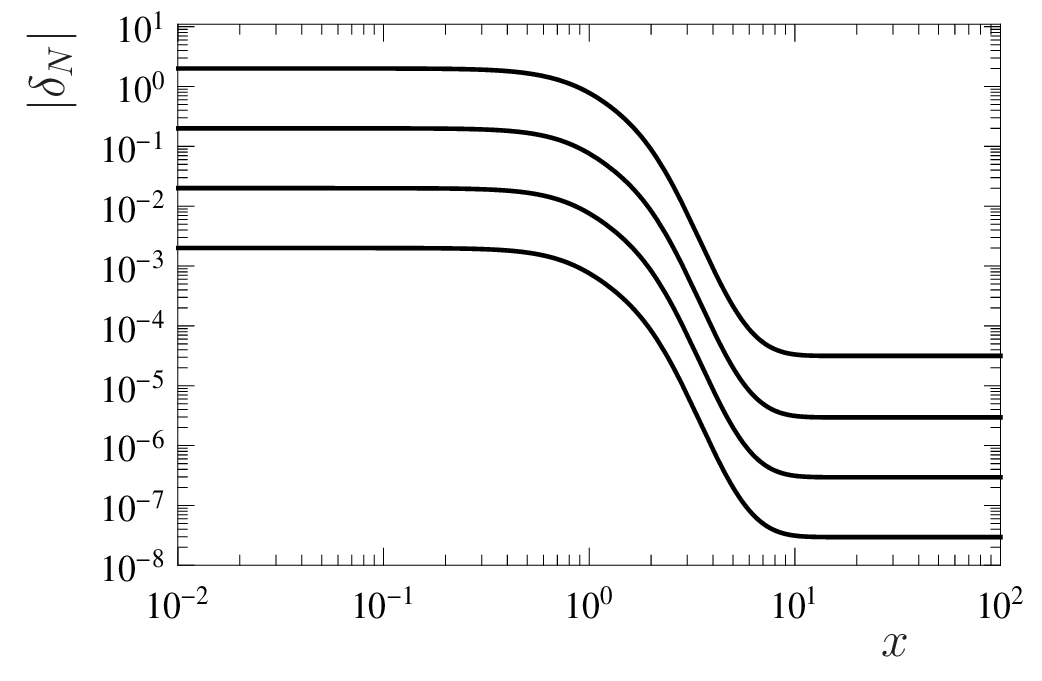} &
\includegraphics[width=0.48\textwidth]{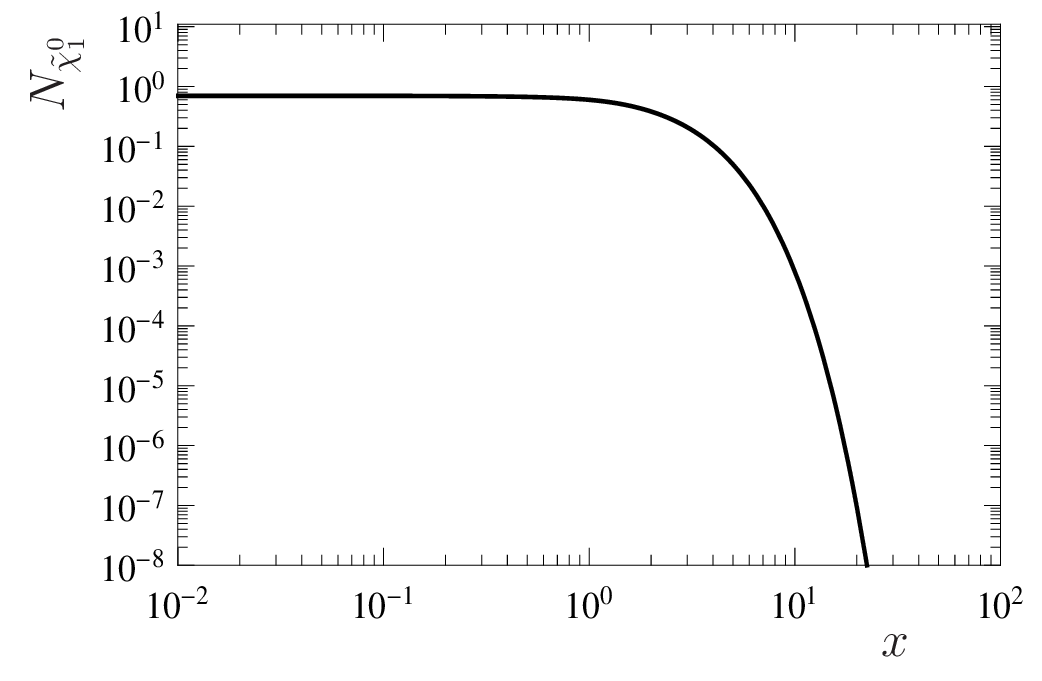}  \\[-0.3cm]
 (a) & (b)
\end{tabular}
\end{center}
\vspace{-5mm}
\caption{(a) Lepton asymmetry $|\delta_N|$ as defined
in \eqn{eq:leptonasymmetry} as a function of $x$
for \cp{}~conserving \brpv{} for all widths set to $5\cdot 10^{-14}$~GeV
for different asymmetries $\delta_N$ at $x=10^{-2}$;
(b) \lsp{} density as a function of $x$ for the same cases.}
\label{fig:realcasenoneutrinodata}
\begin{center}
\includegraphics[width=0.48\textwidth]{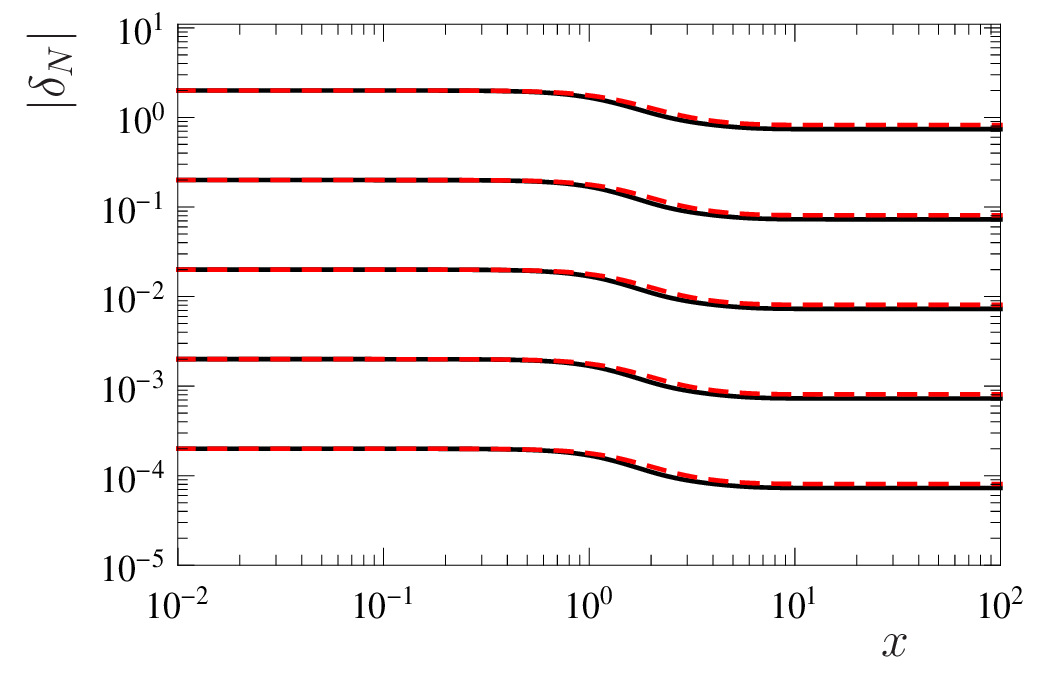}
\vspace{-3mm}
\caption{Lepton asymmetry $|\delta_N|$ as defined
in \eqn{eq:leptonasymmetry} as a function of $x$
for \cp{}~conserving \brpv{} for $P_1$ (black) and $P_2$ (red, dashed)
for different asymmetries $\delta_N$ at $x=10^{-2}$.}
\label{fig:realcaseP1P2}
\end{center}
\end{figure}

In this subsection we discuss the effect of the \lsp{} decays on
the number densities and thus the lepton and baryon asymmetries in the early
universe and we start with the case of \cp{}~conserving \brpv{}.
The small decay rates of the neutralino can have a sizable impact
on lepton asymmetries, which could for example be 
generated at an earlier stage of the universe by the Affleck-Dine 
mechanism \cite{Affleck:1984fy}. If for the moment we ignore neutrino data
and choose all decay widths
slightly larger than the Hubble parameter $\Gamma> H(T=m_{\chiN})$,
a sizable wash-out of initial lepton asymmetries can occur \cite{Campbell:1991at}.
This is explicitly demonstrated in 
\fig{fig:realcasenoneutrinodata}, where all widths are set
to $5\cdot 10^{-14}$~GeV and the annihilation cross sections are taken
from scenario~$P_2$. For a light neutralino $m_{\chiN}<T_c$ initial baryon asymmetries
are only mildly affected, since for $T<T_c$ they are decoupled from the lepton asymmetries
- see \sct{sec:numbaryo}.
For numerical stability we choose all initial particle densities
equal to their equilibrium density with (small)
displacements to establish the shown asymmetries.
The presented effects are independent of the sign of the initial lepton asymmetry $\delta_N$.
\fig{fig:realcasenoneutrinodata}~(b) shows the behaviour of
the \lsp{} number density, which follows the equilibrium
density and thus motivates the neglect of scattering processes.

In accordance to \citeres{Campbell:1992jd,Akeroyd:2003pb} the wash-out of an initial
asymmetry driven by the back-reaction 
of leptons, quarks, neutrinos and their antiparticles to the
\lsp{} and the decays of the \lsp{} itself stays small, if just one of the flavor
final states is suppressed with respect to the others and thus decays
out of equilibrium.
To confirm this statement \fig{fig:realcaseP1P2} shows the small wash-out for different
initial lepton asymmetries for both scenarios $P_1$ and $P_2$ with fulfilled neutrino data.
Initially present lepton and thus also baryon asymmetries are almost conserved, if neutrino
data is explained by the \brpv{} parameters.

\subsection{Baryogenesis via Leptogenesis in the \cp{}~violating \brpv{}}
\label{sec:numbaryo}

In this subsection we discuss the impact of \cp{}~violation by
complex \brpv{} parameters $\epsilon_i$ on the decay widths of the lightest
neutralino and the lepton and baryon asymmetries in the early universe.
Before doing so let us briefly comment on the stringent bounds coming
from the non-observation of electric dipole-moments, 
in particular the one of the electron has to be below $\lesssim 10^{-28}$~ecm
\cite{Baron:2013eja}. As we only consider \cp{}~phases in the $R$-parity
violating parameters, the corresponding effect is small and in case
of the slepton and sneutrino contributions further suppressed by their heavy
masses. The potentially most troublesome are the $\tilde \chi^0_j$-$W$ contributions
which are proportional to Im($ O^W_{Lej} (O^W_{Rej})^*$). Using an expansion
in the $R$-parity violating parameters \cite{Porod:2000hv,Liebler:2011tp} one 
finds that this product is tiny for several reasons: (i) it is proportional
to the $R$-parity violation couplings squared, (ii) it is suppressed by
a factor $Y^{11}_E v_d/$min$(\mu,M_2$) and (iii) it vanishes completely
in case of a pure bino. Numerically we find that the induced electron
dipole-moment is always below $\mathcal{O}(10^{-32}$ ecm) in our examples.

\begin{figure}[htp]
\begin{center}
\begin{tabular}{cc}
\includegraphics[width=0.48\textwidth]{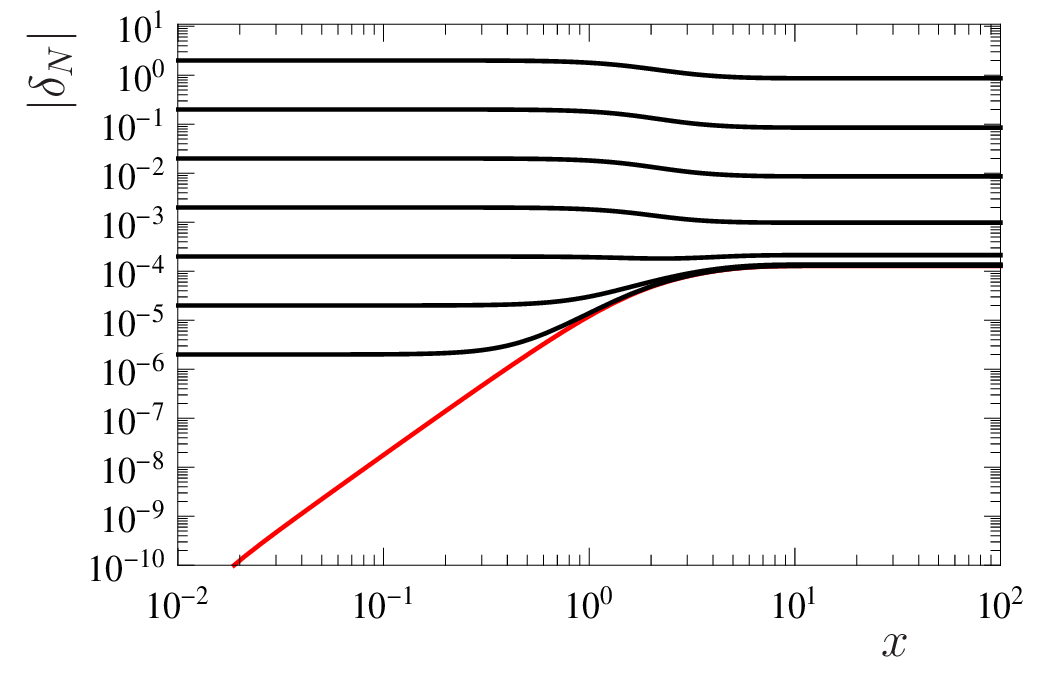} &
\includegraphics[width=0.48\textwidth]{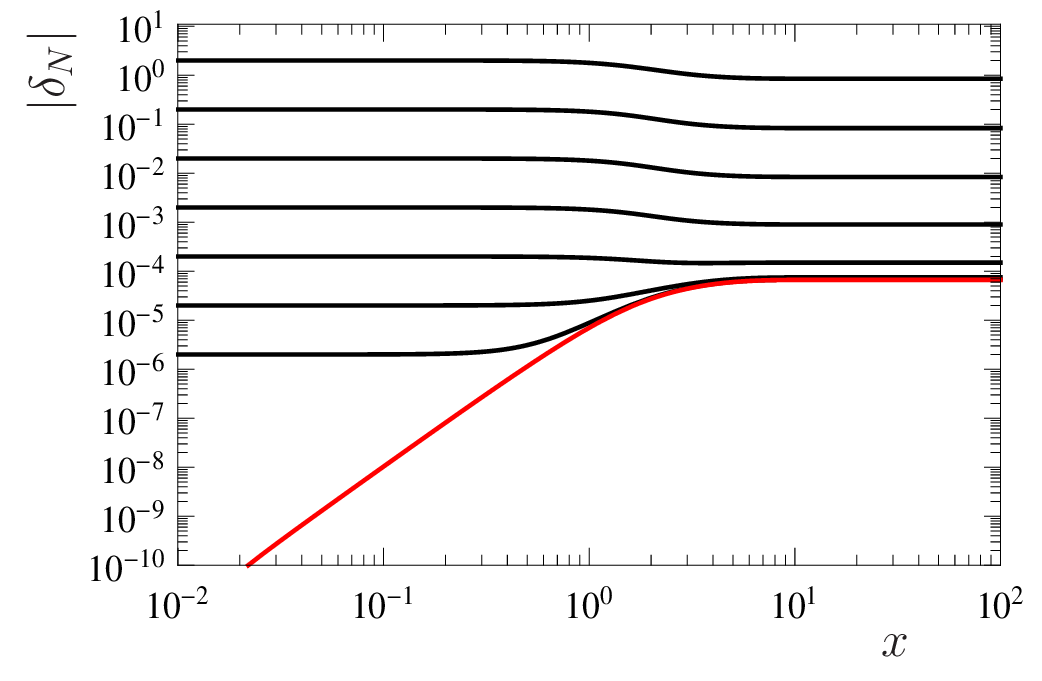}  \\[-0.3cm]
 (a) & (b)
\end{tabular}
\end{center}
\vspace{-5mm}
\caption{Lepton asymmetry $|\delta_N|$ as defined in \eqn{eq:leptonasymmetry}
for (a) $P_1$ with phase $\phi:=\phi_1=\phi_2=\phi_3=5$ degrees;
(b) $P_2$ with phase $\phi:=\phi_1=\phi_2=\phi_3=15$ degrees
in both cases for different initial asymmetries $\delta_N$ at $x=10^{-2}$.}
\label{fig:densitylargeCPasymP1}
\begin{center}
\includegraphics[width=0.48\textwidth]{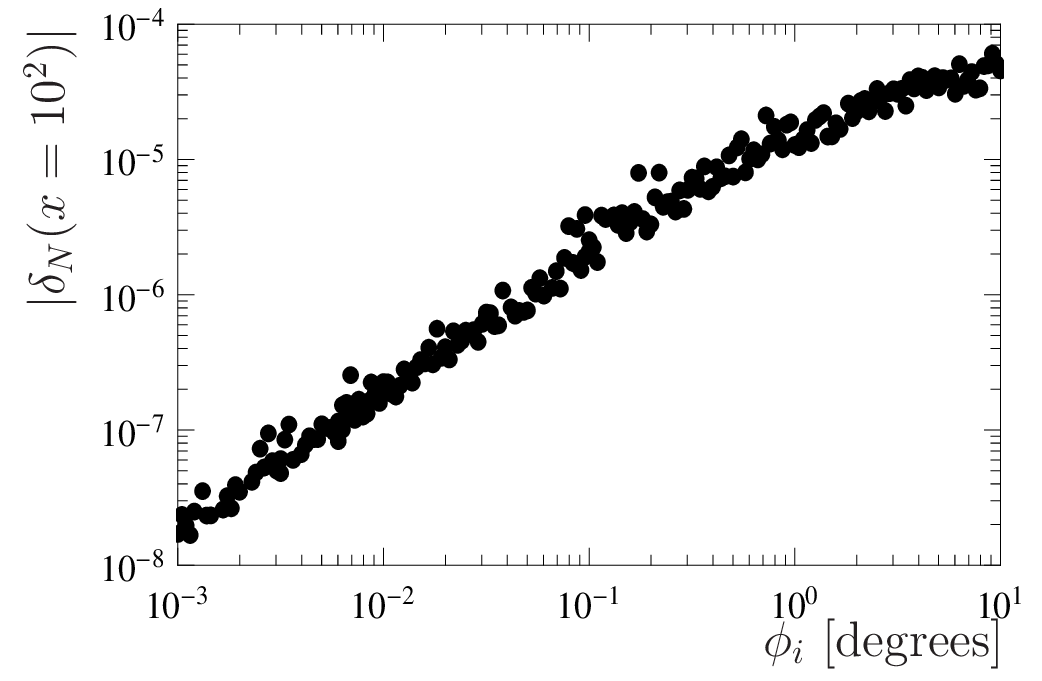}
\vspace{-3mm}
\caption{Resulting lepton asymmetry $|\delta_N|$ at $x=10^2$
as a function of $\phi:=\phi_1=\phi_2=\phi_3$ for $P_2$ for zero
initial asymmetry $\delta_N=0$ at $x=10^{-2}$.}
\label{fig:P2logrun}
\end{center}
\begin{center}
\begin{tabular}{cc}
\includegraphics[width=0.48\textwidth]{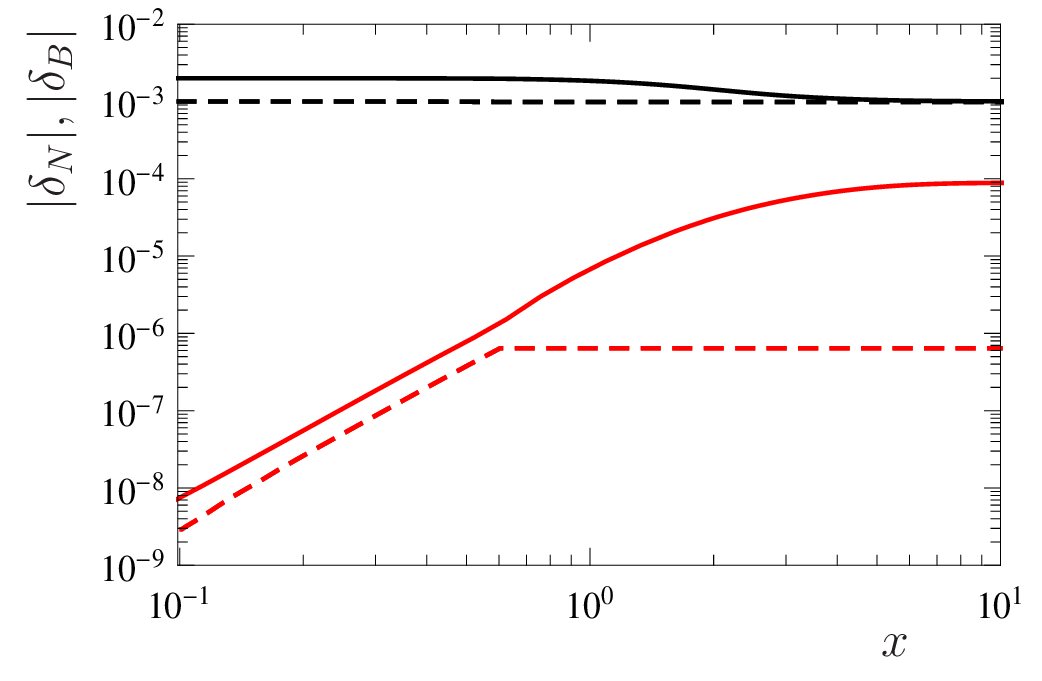} &
\includegraphics[width=0.48\textwidth]{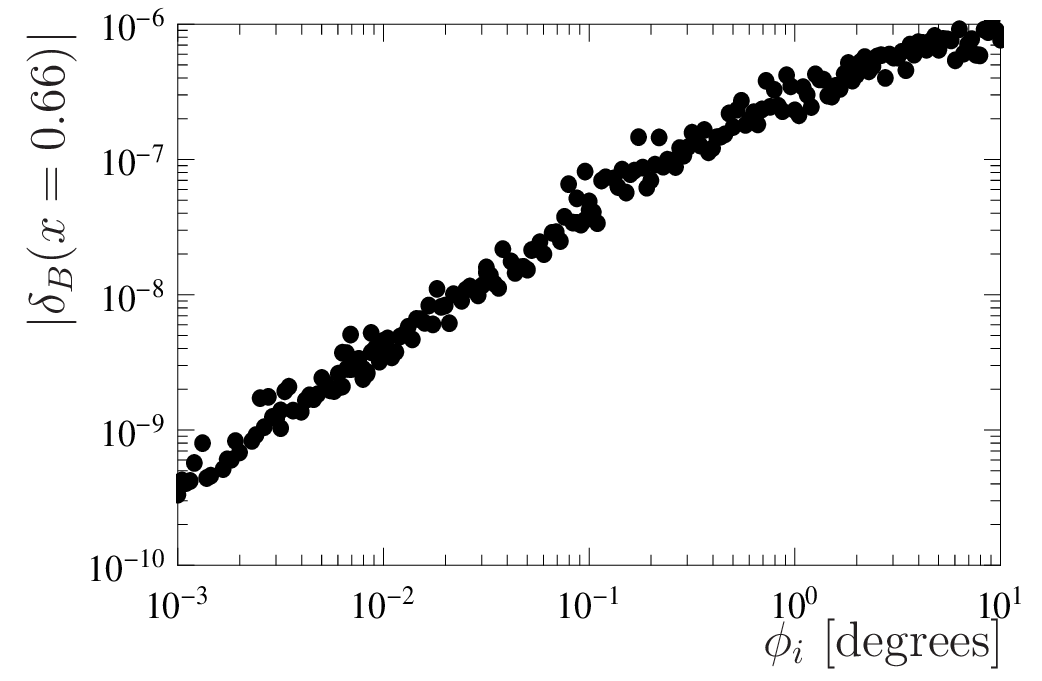}  \\[-0.3cm]
 (a) & (b)
\end{tabular}
\end{center}
\vspace{-5mm}
\caption{(a) Lepton asymmetry $|\delta_N|$ (solid) and baryon asymmetry $|\delta_B|$ (dashed)
for scenario~$P_2$ with phase $\phi:=\phi_1=\phi_2=\phi_3=15$ degrees
for two different initial asymmetries $\delta_N,\delta_B$ at $x=10^{-2}$;
(b) Resulting baryon asymmetry $|\delta_B(x=0.66)|$
as a function of $\phi:=\phi_1=\phi_2=\phi_3$ for $P_2$ for zero
initial asymmetry $\delta_N=\delta_B=0$ at $x=10^{-2}$.}
\label{fig:P2baryo}
\end{figure}

For relatively large phases $\phi_i$ and thus \cp{}~asymmetries up to per-mile 
level the \cp{}~violating contributions can have sizable effects on
lepton asymmetries in the universe.
If the initial lepton asymmetry $\delta_N$ is large, the effect of the \brpv{} induced wash-out dominates -
as discussed in the previous subsection. But
for initial lepton asymmetries being rather small $|\delta_N|<10^{-5}$ the \cp{}~violating contributions
to the \lsp{} decays come into the game. 
They induce a lepton asymmetry of $|\delta_N|\sim 10^{-5}-10^{-3}$, if the complex phases
$\phi_i$ are chosen large $\sim 1$ degree.
Details can be taken from \fig{fig:densitylargeCPasymP1} for scenario $P_1$ choosing
a phase of $5$ degrees and scenario $P_2$ with a phase of $15$ degrees.
\fig{fig:P2logrun} shows the obtained lepton asymmetry in the universe as a function
of the \cp{}~phases $\phi_i$ in the \brpv{} parameters for scenario~$P_2$,
if initially no lepton asymmetry is present $\delta_N(x=10^{-2})=0$.

As pointed out in the previous section $\sin\phi_i\rightarrow - \sin\phi_i$ for all $i\in\lbrace 1,2,3\rbrace$
results in $\delta_\Gamma \rightarrow -\delta_\Gamma$, which induces $\delta_N\rightarrow -\delta_N$,
if no initial asymmetry is present. This statement implies, that for different signs in $\sin\phi_i$
also smaller lepton asymmetries with different signs for
the three generations can be accommodated.
Additionally a cancellation between existing lepton asymmetries and the generated
lepton asymmetries can be arranged.

As discussed in \sct{sec:theobaryo} we add sphaleron transitions to
our Boltzmann equations in order to determine the baryon asymmetry generated
from the lepton asymmetry and vice versa. We therefore start once with initial and once
without initial lepton and
baryon asymmetries and examine how both evolve as a function of the temperature parameterized by $x$.
\fig{fig:P2baryo}~(a) shows the corresponding results for $P_2$ again for a phase
of $\phi:=\phi_1=\phi_2=\phi_3=15$ degrees. For numerical stability we choose
$\delta_B=-\eta(x)\delta_L$ at $x=10^{-2}$ in case of given initial asymmetries.
As expected, the
baryon asymmetry $\delta_B$ freezes at temperatures $x=m_{\chiN}/T_c\approx 0.66$,
whereas the lepton asymmetry $\delta_N$ evolves further driven by the neutralino decays.
Thus, if an initial baryon asymmetry is present, it is hardly affected by
\cp{}~violating decays of light neutralinos. The reason is, that for $x\gtrsim 0.66$, where
the change of the lepton asymmetry is strongest, the
sphaleron process is close to be frozen out.
On the other hand, even in case of no initial baryon asymmetry it can be generated at small $x$. 
We translate \fig{fig:P2logrun} to the corresponding baryon asymmetry
obtained at $x\approx 0.66$ and present the result in \fig{fig:P2baryo}~(b).
Since the baryon asymmetry remains constant for lower temperatures, it yields
$\delta_B(x=10^2)=\delta_B(x=0.66)$.
A baryon asymmetry of order $10^{-10}$ as observed in the universe
can be generated from neutralino
decays having a mass of $m_{\chiN}\sim 100$\,GeV
in \brpv{} in case of rather small complex phases $\phi_i\sim 10^{-3}$ degrees
for the $R$-parity breaking
parameters. The generated lepton asymmetry is approximately two orders
of magnitude larger. Alternatively larger phases are possible, if a cancellation
between various \cp{}~violating contributions occurs.

\subsection{\lsp{} annihilation to \sm{} particles}

\begin{figure}[htp]
\begin{center}
\begin{tabular}{cc}
\includegraphics[width=0.48\textwidth]{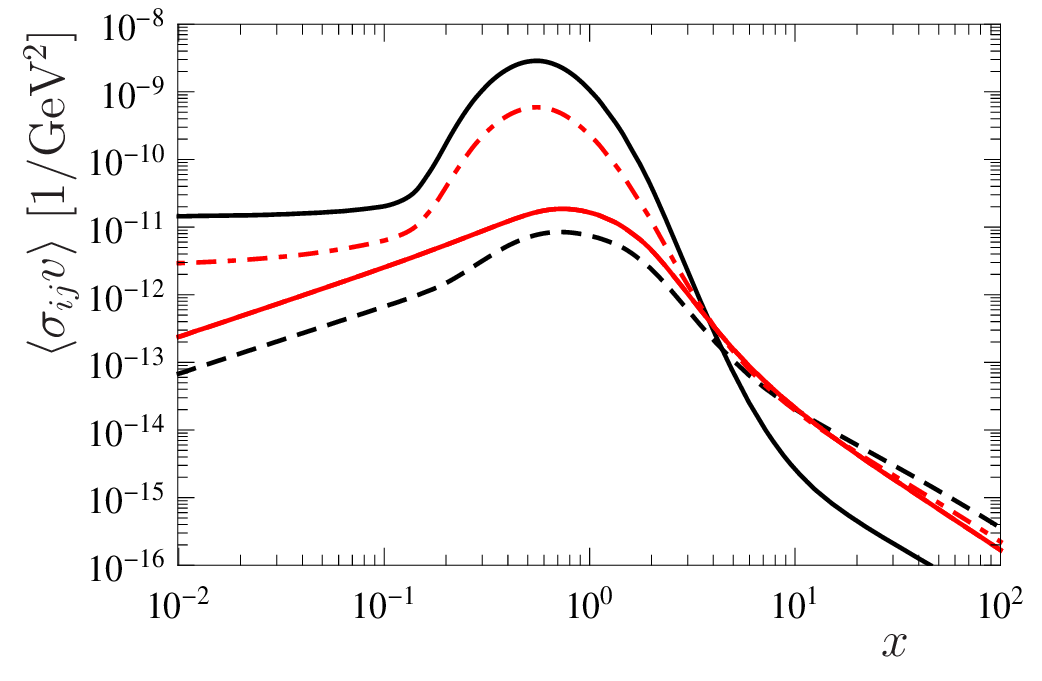} &
\includegraphics[width=0.48\textwidth]{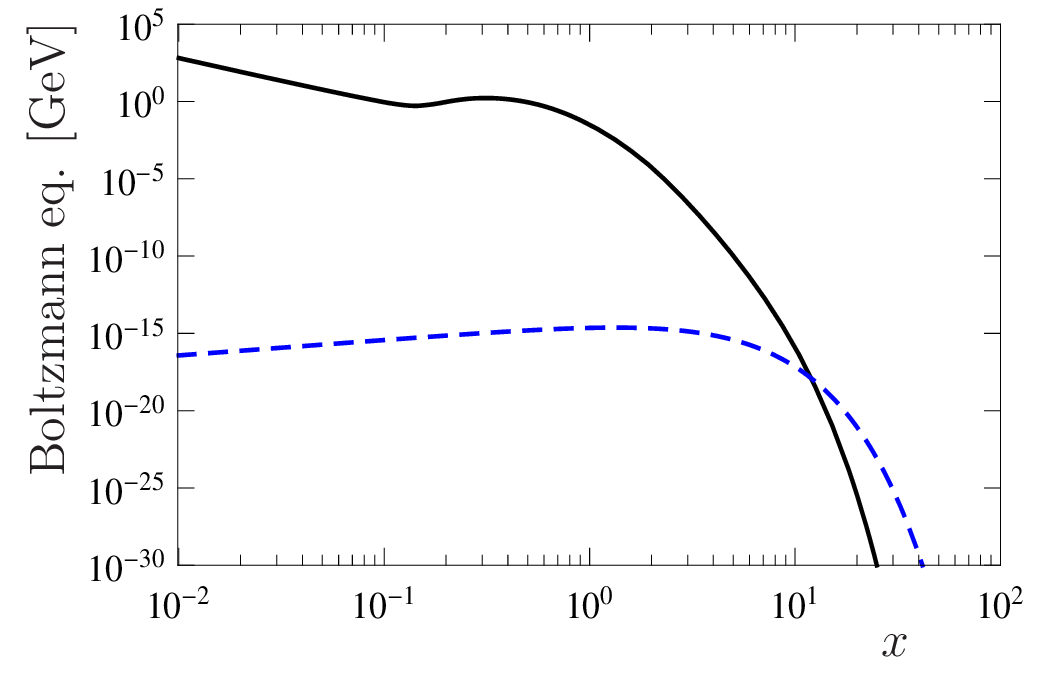}  \\[-0.3cm]
 (a) & (b)
\end{tabular}
\end{center}
\vspace{-5mm}
\caption{(a) Cross section $\langle \sigma_{ij} v\rangle$ in $1/$GeV$^2$
as a function of $x$ for $P_2$ for different final states:
$ij=d\bar{d}/s\bar{s}/b\bar{b}$ (black), $u\bar{u}/c\bar{c}/t\bar{t}$ (black, dashed),
$\tau^+\tau^-$ (red, dot-dashed), $e^+e^-/\mu^+\mu^-$ (red);
(b) $K_1/K_2 N^{eq} \Gamma$ in GeV with $\Gamma=10^{-14}$~GeV (blue, dashed) and
$\hat{\sigma}(\chiN\chiN\rightarrow b\bar{b})N^{eq,2}$ in GeV (black) as a function of $x$.}
\label{fig:AXS}
\begin{center}
\begin{tabular}{cc}
\includegraphics[width=0.48\textwidth]{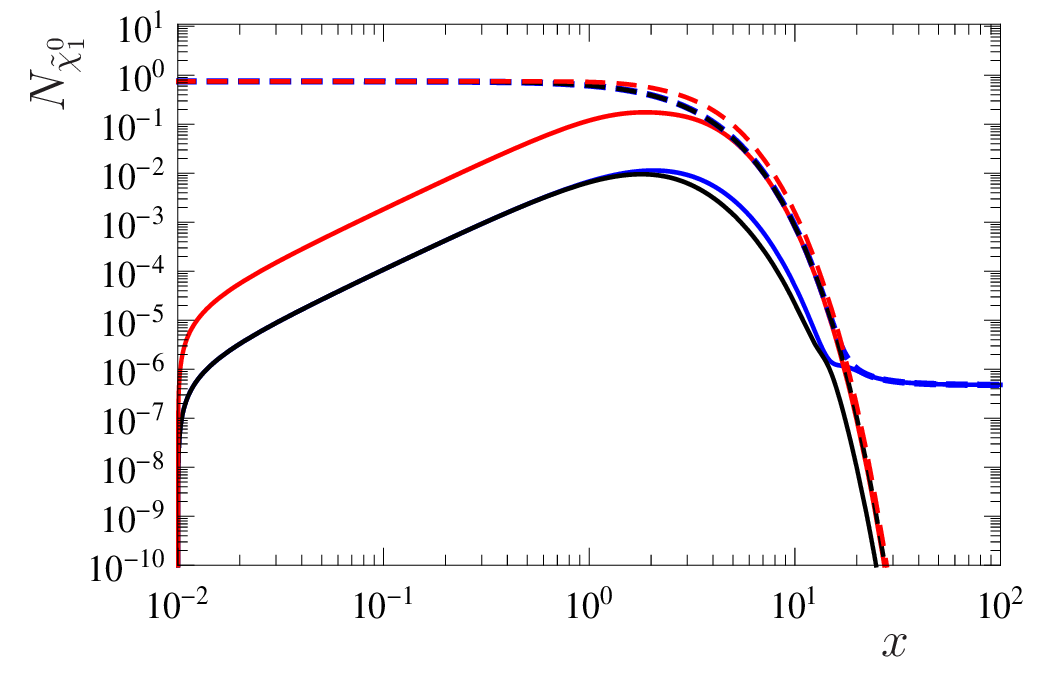} &
\includegraphics[width=0.48\textwidth]{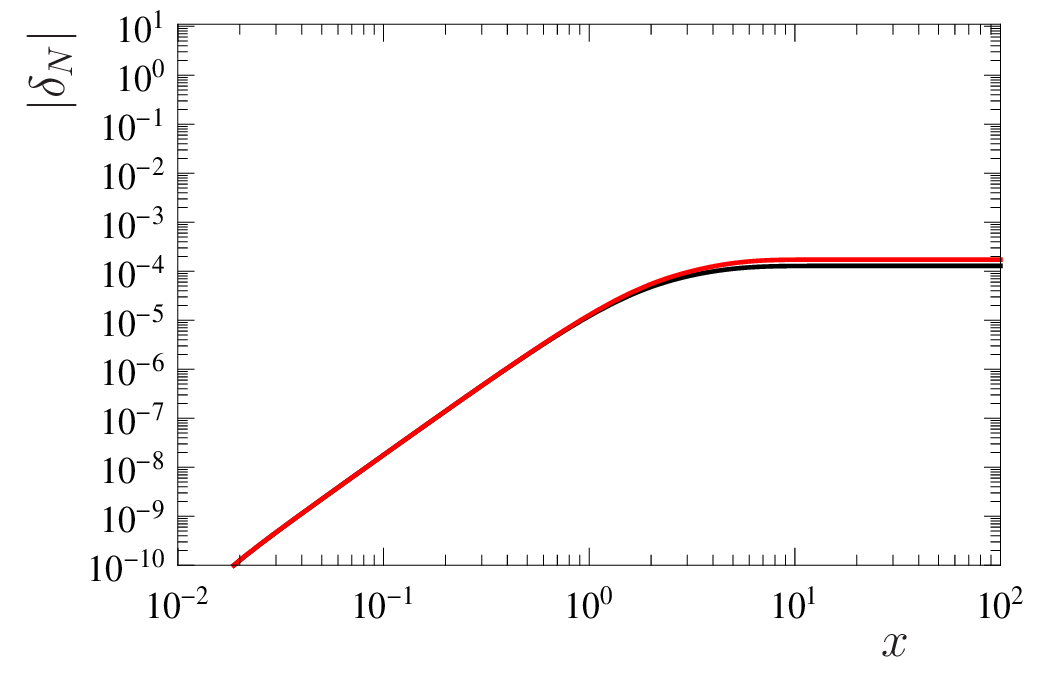}  \\[-0.3cm]
 (a) & (b)
\end{tabular}
\end{center}
\vspace{-5mm}
\caption{(a) \lsp{} number density $|N_{\chiN}-N^{eq}_{\chiN}|$ (solid)
and $N_{\chiN}$ (dashed) as a function of $x$
for only annihilation (blue), only decays (red), annihilation and decays (black) for $P_1$
with phase $\phi_i=5$ degrees;
(b) Evolution of $\delta_N$ as defined
in \eqn{eq:leptonasymmetry} as a function of $x$ for only decays (red),
annihilation and decays (black) for the same case.}
\label{fig:ANNIP1}
\end{figure}

As it was pointed out in \citere{Hambye:2001eu} $R$-parity conserving
scattering processes, which lead to an annihilation of the \lsp{},
can impact on the densities of the \lsp{} and the \sm{} particles.
Therefore all our discussion
included the $R$-parity conserving annihilation processes.
Within this section we want to discuss their impact in
more detail, neglecting sphaleron transitions for simplicity.
For scenario~$P_2$ \fig{fig:AXS}~(a) shows the
thermally averaged cross sections $\langle \sigma_{ij} v \rangle$
for different final states in $1/$GeV$^2$ as obtained
by {\tt micrOMEGAs}. \fig{fig:AXS}~(b)
presents for comparison  
$\hat{\sigma}(\chiN\chiN\rightarrow b\bar{b})N^{eq,2}$
with $N^{eq}$ being the neutralino equilibrium density
versus $\mathrm{K}_1/\mathrm{K}_2 N^{eq} \Gamma$ with
$\Gamma=10^{-14}$~GeV and the ratio $\mathrm{K}_1/\mathrm{K}_2$
of modified Bessel functions
and thus reflects the right-hand side terms entering the
Boltzmann equations \eqn{eq:neutralinodensevol2}.
The annihilation processes dominate the evolution of the
Boltzmann equations for most temperatures, their relative
importance drops below the decay processes only for very
low temperatures $T< m_{\chiN}$.

The impact of the annihilation versus the decay processes in the
Boltzmann equations is further elaborated in
\fig{fig:ANNIP1} for scenario $P_1$. Scenario $P_2$ is visually
indistinguishable from $P_1$.
The left \fig{fig:ANNIP1}~(a) shows the neutralino density $N_{\chiN}$ and its difference
to the corresponding equilibrium value $|N_{\chiN}-N^{eq}_{\chiN}|$ for 
three different scenarios: In case of only annihilation processes,
which corresponds to a scenario of the \mssm{} with $R$-parity conservation,
the well-known freeze-out of the lightest \susy{} particle yielding a
constant neutralino density is observed at low temperatures.
If just \cp{}~violating decays of the neutralinos are included, the
neutralino density vanishes for large $x$. In case of annihilation
and decay processes the neutralino density closely follows the
case of only annihilation processes, but all neutralinos tend to decay
at low temperatures.

The right \fig{fig:ANNIP1}~(b) presents the evolving lepton asymmetry, if no
initial lepton asymmetry $\delta_N=0$ at $x=10^{-2}$ is assumed.
A difference between just decay processes
and decay as well as annihilation processes
can be observed. However, the order of magnitude of the generated
lepton asymmetry remains unchanged. The inclusion of sphaleron
transitions will distort the observations only minimally.

\section{Conclusion}
\label{sec:conclusions}

We examined the effects of \lsp{} decays, the \lsp{}
being a light neutralino $m_{\chiN}\sim 100$\,GeV,
in bilinear $R$-parity violation (\brpv) with complex \brpv{} parameters on
the lepton and baryon asymmetries in the early universe.
We presented a description of the neutralino sector at \nlo{}
 for complex \brpv{} parameters
and calculated the \lsp{} decays
at \nlo{}. In this way we get both, the total width as well as the induced
\cp{}~asymmetries between leptonic and antileptonic final states.
With respect to the evolution of number densities in the
early universe our discussion includes - apart from the mentioned
\lsp{} decays and their inverse counterparts - the \lsp{} annihilation
to \sm{} particles, which we assume to be very close to the minimal
supersymmetric standard model. In order to describe the transition between
lepton and baryon asymmetries we add a simple description of
sphaleron processes, which however get frozen out at the mass scale
of the neutralino if its mass is below the electroweak
phase transition.

Our conclusion is two-sided: Initial lepton and baryon asymmetries
are preserved by the \lsp{} decay, if neutrino data is described
correctly by the \brpv{} parameters. 
On the other hand instead, lepton and baryon asymmetries can also
be generated in the complex \brpv{} model, the latter being in accordance
to the observation in our universe. Clearly, both statements hold for
different values of the \cp{}~violating phases in case of
a light neutralino \lsp{}. Last but not least
we note that in both regimes the via those couplings
induced electric dipole-moment of the
electron is well below the sensitivity of present and future experiments.

\paragraph{Acknowledgments}

We thank M.~Hirsch and Th.~Konstandin for discussions and
 U.~Langenfeld for collaboration in the early stage of this project. 
This work is supported by the DFG, project no.\ PO-1337/2-1, and
by the research training group GRK 1147. S.~L.~acknowledges support by
the DFG through the SFB~676 ``Particles, Strings and the Early Universe''.
S.~L.~was also supported by the DFG grant HA 2990/5-1.

\newpage
\begin{appendix}

\section{Formulas: Boltzmann equations}
\label{sec:formulas}

In this section we present the full set of Boltzmann equations for the
neutralino and the final state particles resulting from the gauge bosons/Higgs
decays namely leptons, neutrinos (antineutrinos) and quarks (antiquarks).
To be brief we just add the decay mode $h\rightarrow q\bar q$ for the Higgs boson.
Moreover to shorten the subsequent formulas, we leave the distinction of $q_1$ and $q_2$
to the reader. It is clear that e.g. the decays of the $W$ boson involve two
different quark types, whereas the decays of the $Z$ boson result in 
identical quark types $q_i\bar q_i$. Accordingly \eqn{eq:q} and \eqn{eq:qbar}
have to be doubled for both quarks $q_1$ and $q_2$. For details
with respect to the sphaleron transitions we refer to \sct{sec:theobaryo}.
To shorten our notation we write $\Gamma,\Br(X|YZ):=\Gamma,\Br(X\rightarrow YZ)$ and
$\hat{\sigma}(X X|Y Z):=\hat{\sigma}(X X \rightarrow Y Z)$. 
We also define $\tN_X$ to be the ratio of the number density of the particle
$X\in \lbrace \ell^\pm_{i},\nu,\bar\nu,q,\bar q\rbrace$ and its value
in the thermal equilibrium:
\begin{align}
 \tN_X = \frac{N_X}{N^{eq}_X} 
\end{align}

{\allowdisplaybreaks
\begin{align}
\begin{split}
xH\frac{dN_{\chiN}}{dx} &=  -\frac{\mathrm{K}_1(x)}{\mathrm{K}_2(x)}\underset{i,j}{\sum}\underset{q,\bar{q}}{\sum}
\Bigl[ N_{\chiN}\Bigl(\Gamma(\chiN|\ell^+_{i}W^-)+\Gamma(\chiN|\ell^-_{i}W^+)\Bigr)
 \Br(W^\pm|\ell^\pm_j \nu (\bar{\nu})_j) \\ & -   
\tN_{\ell^+_{i}} \tN_{\ell^-_{j}} \tN_{\bar{\nu}_j}N_{\chiN}^{eq}
\Gamma(\ell^+_{i} W^-|\chiN) \Br(W^-|\ell^- _{j} \bar{\nu}_j) \\ & -
\tN_{\ell^-_{i}} \tN_{\ell^+_{j}}\tN_{\nu_j}N_{\chiN}^{eq} 
\Gamma(\ell^-_{i} W^+|\chiN) \Br(W^+|\ell^+ _{j} \nu_j) \\ & + 
N_{\chiN}\Bigl(\Gamma(\chiN|\ell^- _{i}W^+) +
\Gamma(\chiN|\ell^+ _{i}W^-)\Bigr) \Br(W^\pm |q \bar q)
\\ &
-  \tN_{q}\tN_{\bar q} N_{\chiN}^{eq} 
\Bigl(\tN_{\ell^-_i}\Gamma(\ell^-_{i} W^+|\chiN)+\tN_{\ell^+_i}\Gamma(l^+_{i} W^-|\chiN)\Bigr) \Br(W^\pm |q \bar q)  \\ & +
N_{\chiN} \Bigl(\Gamma(\chiN| Z \nu_i)+\Gamma(\chiN|Z \bar{\nu}_i)\Bigr) 
\Br(Z |\ell^- _{j} \ell^+ _{j}) \\ & - 
\tN_{\ell^-_j}\tN_{\ell^+_j} N_{\chiN}^{eq} \Bigl(\tN_{\nu_i} \Gamma (Z \nu_i| \chiN)
+ \tN_{\bar{\nu}_i} \Gamma (Z \bar{\nu}_i| \chiN)\Bigr) \Br(Z |\ell^- _{j}\ell^+_{j})
\\ & +
 N_{\chiN} \Bigl(\Gamma(\chiN| Z \nu_i)+\Gamma(\chiN| Z \bar{\nu}_i)\Bigr) \Br(Z | q \bar q)
\\ & -
 \tN_{q}\tN_{\bar q} N_{\chiN}^{eq}
 \Bigl( \tN_{\nu_i} \Gamma(Z \nu_i| \chiN) + \tN_{\bar{\nu}_i} \Gamma (Z \bar{\nu}_i| \chiN) \Bigr)\Br(Z | q \bar q)
\\ & +
 N_{\chiN} \Bigl(\Gamma(\chiN| h \nu_i)+\Gamma(\chiN| h \bar{\nu}_i)\Bigr) \Br(h | q \bar q)
\\ & -
 \tN_{q}\tN_{\bar q} N_{\chiN}^{eq}
 \Bigl( \tN_{\nu_i} \Gamma(h \nu_i| \chiN) + \tN_{\bar{\nu}_i} \Gamma (h \bar{\nu}_i| \chiN) \Bigr)\Br(h | q \bar q)
\\ &
+ N_{\chiN} \Gamma(\chiN| Z \nu_i) \Br(Z | \nu_j \bar{\nu}_j ) 
- \tN_{\nu_i} \tN_{\nu_j}\tN_{\bar{\nu}_j}N_{\chiN}^{eq} \Gamma (Z \nu_i| \chiN)
\Br(Z | \nu_j \bar{\nu}_j)
\\ &
+ N_{\chiN} \Gamma(\chiN| Z \bar{\nu}_i) \Br(Z | \nu_j \bar{\nu}_j ) 
- \tN_{\bar{\nu}_i} \tN_{\nu_j}\tN_{\bar{\nu}_j}N_{\chiN}^{eq} \Gamma (Z \bar{\nu}_i| \chiN)
\Br(Z | \nu_j \bar{\nu}_j) \Bigr]
 \\ & - 
\underset{i}{\sum}\underset{q,\bar{q}}{\sum} \Bigl[
\hat{\sigma}(\chiN \chiN|\ell^+_i \ell^-_i) (N^2_{\chiN}-\tN_{\ell^+_i}\tN_{\ell^-_i} {N_{\chiN}^{eq}}^2) \\ & +
\hat{\sigma}(\chiN \chiN|\nu_i \bar \nu_i) (N^2_{\chiN}-\tN_{\nu_i}\tN_{\bar \nu_i} {N_{\chiN}^{eq}}^2) +
 \hat{\sigma}(\chiN \chiN|q \bar q) (N^2_{\chiN}-\tN_{q}\tN_{\bar q} {N_{\chiN}^{eq}}^2) 
 \Bigr]
\end{split}
\end{align}
}

{\allowdisplaybreaks
\begin{align}
\begin{split}
xH\frac{dN_{\nu_j}}{dx} &= \frac{\mathrm{K}_1(x)}{\mathrm{K}_2(x)}\underset{i}{\sum}\underset{q,\bar{q}}{\sum}
\Bigl[ N_{\chiN}\Gamma(\chiN|\ell^- _{i}W^+) \Br(W^+|\ell^+ _{j} \nu_j)  \\ & -
\tN_{\ell^-_{i}}\tN_{\ell^+_{j}}\tN_{\nu_j}N_{\chiN}^{eq} 
\Gamma(\ell^-_{i} W^+|\chiN) \Br(W^+|\ell^+ _{j} \nu_j) \\ & +
N_{\chiN} \Gamma(\chiN| Z \nu_j) \Bigl(\Br(Z |\ell^- _{i}\ell^+_{i})
+ \Br(Z | \nu_i \bar{\nu}_i) + 
\Br(Z | q  \bar q) \Bigr) \\ & -
\tN_{\nu_j}\tN_{\ell^-_i}\tN_{\ell^+_i} N_{\chiN}^{eq} \Gamma (Z \nu_j| \chiN)
\Br(Z |\ell^- _{i}\ell^+ _{i} ) -
 \tN_{\nu_j} \tN_{q}\tN_{\bar q} N_{\chiN}^{eq}   
\Gamma (Z \nu_j| \chiN)  \Br(Z | q \bar q) 
\\ & +
N_{\chiN} \Gamma(\chiN| h \nu_j) 
\Br(h | q  \bar q) -
 \tN_{\nu_j} \tN_{q}\tN_{\bar q} N_{\chiN}^{eq}   
\Gamma (h \nu_j| \chiN)  \Br(h | q \bar q) 
\\ & -
\tN_{\nu_j} \tN_{\nu_i}\tN_{\bar{\nu}_i} N_{\chiN}^{eq} \Gamma (Z \nu_j| \chiN)
\Br(Z | \nu_i \bar{\nu}_i) \\ & +
N_{\chiN} \Gamma(\chiN| Z \bar{\nu}_j) \Br(Z | \nu_i \bar{\nu}_i)
- \tN_{\bar{\nu}_j} \tN_{\nu_i}\tN_{\bar{\nu}_i} N_{\chiN}^{eq} \Gamma (Z \bar{\nu}_j| \chiN)
\Br(Z | \nu_i \bar{\nu}_i) \\ &
+ N_{\chiN} \Gamma(\chiN| Z \nu_i) \Br(Z | \nu_j \bar{\nu}_j)
- \tN_{\nu_i} \tN_{\nu_j}\tN_{\bar{\nu}_j} N_{\chiN}^{eq} \Gamma (Z \bar{\nu}_j| \chiN)
\Br(Z | \nu_i \bar{\nu}_i)
\Bigr] \\ & +\frac{1}{2}\hat{\sigma}(\chiN \chiN|\nu_j \bar \nu_j) (N^2_{\chiN}-\tN_{\nu_j}\tN_{\bar \nu_j} {N_{\chiN}^{eq}}^2) 
+\frac{\gamma(x)}{12}\Bigl[\delta_B+\eta(x)\delta_L\Bigr]
\end{split}
\end{align}
}

{\allowdisplaybreaks
\begin{align}
\begin{split}
xH\frac{dN_{\bar{\nu}_j}}{dx} &= \frac{\mathrm{K}_1(x)}{\mathrm{K}_2(x)}\underset{i}{\sum}\underset{q,\bar{q}}{\sum}
\Bigl[ N_{\chiN}\Gamma(\chiN|\ell^+ _{i}W^-) \Br(W^-|\ell^- _{j} \bar{\nu}_j)  \\ & -
\tN_{\ell^+_{i}}\tN_{\ell^-_{j}}\tN_{\bar{\nu}_j}N_{\chiN}^{eq} 
\Gamma(\ell^+_{i} W^-|\chiN) \Br(W^-|\ell^- _{j} \bar{\nu}_j) \\ & +
N_{\chiN} \Gamma(\chiN| Z \bar{\nu}_j) \Bigl(\Br(Z |\ell^- _{i}\ell^+_{i})+ 
\Br(Z | \nu_i \bar{\nu}_i) + \Br(Z | q \bar q) \Bigr) \\ & -
\tN_{\bar{\nu}_j}\tN_{\ell^-_i}\tN_{\ell^+_i} N_{\chiN}^{eq} \Gamma (Z \bar{\nu}_j| \chiN)
\Br(Z |\ell^- _{i}\ell^+ _{i}) - 
\tN_{\bar{\nu}_j} \tN_{q}\tN_{\bar q} N_{\chiN}^{eq}\Gamma (Z \bar{\nu}_j| \chiN) \Br(Z | q \bar q) 
\\ & +
N_{\chiN} \Gamma(\chiN| h \bar{\nu}_j) \Br(h | q \bar q) - 
\tN_{\bar{\nu}_j} \tN_{q}\tN_{\bar q} N_{\chiN}^{eq}\Gamma (h \bar{\nu}_j| \chiN) \Br(h | q \bar q)
\\ & -
\tN_{\bar{\nu}_j} \tN_{\nu_i}\tN_{\bar{\nu}_i} N_{\chiN}^{eq} \Gamma (Z \bar{\nu}_j| \chiN)
\Br(Z | \bar{\nu}_i \bar{\nu}_i) \\ & +
N_{\chiN} \Gamma(\chiN| Z \nu_j) \Br(Z | \nu_i \bar{\nu}_i)
- \tN_{\nu_j} \tN_{\nu_i}\tN_{\bar{\nu}_i} N_{\chiN}^{eq} \Gamma (Z \nu_j| \chiN)
\Br(Z | \nu_i \bar{\nu}_i) \\ &
+ N_{\chiN} \Gamma(\chiN| Z \bar{\nu}_i) \Br(Z | \nu_j \bar{\nu}_j)
- \tN_{\bar{\nu}_i} \tN_{\nu_j}\tN_{\bar{\nu}_j} N_{\chiN}^{eq} \Gamma (Z \bar{\nu}_i| \chiN)
\Br(Z | \nu_i \bar{\nu}_i) 
\Bigr]\\ & +\frac{1}{2}\hat{\sigma}(\chiN \chiN|\nu_j \bar \nu_j) (N^2_{\chiN}-\tN_{\nu_j}\tN_{\bar \nu_j} {N_{\chiN}^{eq}}^2) 
-\frac{\gamma(x)}{12}\Bigl[\delta_B+\eta(x)\delta_L\Bigr]
\end{split}
\end{align}
}

{\allowdisplaybreaks
\begin{align}
\label{eq:q}
\begin{split}
xH\frac{dN_{q}}{dx}  &= \frac{\mathrm{K}_1(x)}{\mathrm{K}_2(x)}\underset{i}{\sum} \underset{\bar{q}}{\sum} 
\Bigl[ N_{\chiN}\Gamma(\chiN|\ell^- _{i}W^+)\Br(W^+| q \bar q)  \\ & -
\tN_{\ell^-_i} \tN_{q}  \tN_{\bar q} N_{\chiN}^{eq}\Gamma(\ell^-_{i} W^+|\chiN) 
\Br(W^+| q \bar q) + 
N_{\chiN} \Br(Z | q  \bar q) \Bigl(\Gamma(\chiN| Z \nu_i) + \Gamma(\chiN| Z \bar{\nu}_i) \Bigr) \\ & - 
\tN_{q} \tN_{\bar q} N_{\chiN}^{eq} \Br(Z | q \bar q) 
\Bigl(\tN_{\nu_i}\Gamma (Z \nu_i| \chiN) + \tN_{\bar{\nu}_i} \Gamma (Z \bar{\nu}_i| \chiN)  \Bigr) \\ &
+N_{\chiN} \Br(h | q  \bar q) \Bigl(\Gamma(\chiN| h \nu_i) + \Gamma(\chiN| h \bar{\nu}_i) \Bigr) \\ & - 
\tN_{q} \tN_{\bar q} N_{\chiN}^{eq} \Br(h | q \bar q) 
\Bigl(\tN_{\nu_i}\Gamma (h \nu_i| \chiN) + \tN_{\bar{\nu}_i} \Gamma (h \bar{\nu}_i| \chiN)  \Bigr) \Bigr]
\\ & 
+\underset{\bar q}{\sum} \Bigl[\frac{1}{2}\hat{\sigma}(\chiN \chiN|q \bar q) (N^2_{\chiN}-\tN_{q}\tN_{\bar q} {N_{\chiN}^{eq}}^2) \Bigr] 
+\frac{\gamma(x)}{4}\Bigl[\delta_B+\eta(x)\delta_L\Bigr]
\end{split}
\end{align}
}

{\allowdisplaybreaks
\begin{align}
\label{eq:qbar}
\begin{split}
 xH\frac{dN_{\bar q}}{dx}  &=  \frac{\mathrm{K}_1(x)}{\mathrm{K}_2(x)}\underset{i}{\sum} \underset{q}{\sum} 
\Bigl[ N_{\chiN}\Gamma(\chiN|\ell^+ _{i}W^-)\Br(W^-| q \bar q)  \\ & -
\tN_{\ell^+_i} \tN_{q}  \tN_{\bar q} N_{\chiN}^{eq}\Gamma(\ell^+_{i} W^-|\chiN) 
\Br(W^-| q \bar q) + 
N_{\chiN} \Br(Z | q  \bar q) \Bigl(\Gamma(\chiN| Z \nu_i) + \Gamma(\chiN| Z \bar{\nu}_i) \Bigr) \\ & - 
\tN_{q} \tN_{\bar q} N_{\chiN}^{eq} \Br(Z | q \bar q) 
\Bigl(\tN_{\nu_i}\Gamma (Z \nu_i| \chiN) + \tN_{\bar{\nu}_i} \Gamma (Z \bar{\nu}_i| \chiN)  \Bigr) \\ &
+ N_{\chiN} \Br(h | q  \bar q) \Bigl(\Gamma(\chiN| h \nu_i) + \Gamma(\chiN| h \bar{\nu}_i) \Bigr) \\ & - 
\tN_{q} \tN_{\bar q} N_{\chiN}^{eq} \Br(h | q \bar q) 
\Bigl(\tN_{\nu_i}\Gamma (h \nu_i| \chiN) + \tN_{\bar{\nu}_i} \Gamma (h \bar{\nu}_i| \chiN)  \Bigr) \Bigr]
\\ & 
+\underset{ q}{\sum} \Bigl[\frac{1}{2}\hat{\sigma}(\chiN \chiN|q \bar q) (N^2_{\chiN}-\tN_{q}\tN_{\bar q} {N_{\chiN}^{eq}}^2) \Bigr] 
-\frac{\gamma(x)}{4}\Bigl[\delta_B+\eta(x)\delta_L\Bigr]
\end{split}
\end{align}
}

{\allowdisplaybreaks
\begin{align}
\begin{split}
xH\frac{dN_{\ell^-_{i}}}{dx} &=  \frac{\mathrm{K}_1(x)}{\mathrm{K}_2(x)}\underset{j}{\sum}\underset{q,\bar{q}}{\sum} 
\Bigl[ N_{\chiN} \Gamma(\chiN|\ell^- _{i}W^+) \Br(W^+|\ell^+_j \nu_j) \\ & -
\tN_{\ell^-_{i}}\tN_{\ell ^+_{j}}\tN_{\nu_j}N_{\chiN}^{eq} 
\Gamma(\ell^-_{i} W^+|\chiN) \Br(W^+|\ell^+ _{j} \nu_j) \\& +
N_{\chiN}\Gamma(\chiN|\ell^- _{i}W^+)\Br(W^+| q \bar q) - 
\tN_{\ell^-_i} \tN_{q} \tN_{\bar q} N_{\chiN}^{eq}\Gamma(\ell^-_{i} W^+|\chiN) 
\Br(W^+| q \bar q) \\ & + 
N_{\chiN} \Gamma(\chiN|\ell^+_{j}W^-) \Br(W^-|\ell^-_i \bar{\nu}_i) -
\tN_{l^+_{j}}\tN_{\ell^-_{i}}\tN_{\bar{\nu}_i}N_{\chiN}^{eq} 
\Gamma(l^+_{j} W^-|\chiN) \Br(W^-|\ell^- _{i} \bar{\nu}_i) \\& +
 N_{\chiN} \Gamma(\chiN| Z \nu_j) 
\Br(Z |\ell^- _{i} \ell^+ _{i})   - 
\tN_{\nu_j}\tN_{\ell^-_i}\tN_{\ell^+_i} N_{\chiN}^{eq} \Gamma (Z \nu_j| \chiN)
\Br(Z |\ell^- _{i}\ell^+_{i}) \\ & + 
N_{\chiN} \Gamma(\chiN| Z \bar{\nu}_j) 
\Br(Z |\ell^- _{i} \ell^+ _{i})  - 
\tN_{\bar{\nu}_j}\tN_{\ell^-_i}\tN_{\ell^+_i} N_{\chiN}^{eq} \Gamma (Z \bar{\nu}_j| \chiN)
\Br(Z |\ell^- _{i}\ell^+_{i})
\Bigr] \\ & +\hat{\sigma}(\chiN \chiN|\ell^+_i \ell^-_i) (N^2_{\chiN}-\tN_{\ell^+_i}\tN_{\ell^-_i} {N_{\chiN}^{eq}}^2) 
+\frac{\gamma(x)}{12}\Bigl[\delta_B+\eta(x)\delta_L\Bigr]
\end{split}
\end{align}
}

{\allowdisplaybreaks
\begin{align}
\begin{split}
 xH\frac{dN_{\ell^+_{i}}}{dx}= & \frac{\mathrm{K}_1(x)}{\mathrm{K}_2(x)}\underset{j}{\sum}\underset{q,\bar{q}}{\sum} 
\Bigl[ N_{\chiN} \Gamma(\chiN|\ell^+ _{i}W^-) \Br(W^-|\ell^-_j \bar{\nu}_j) \\ & -
\tN_{l^+_{i}}\tN_{\ell^-_{j}}\tN_{\bar{\nu}_j}N_{\chiN}^{eq} 
\Gamma(l^+_{i} W^-|\chiN) \Br(W^-\to\ell^- _{j} \bar{\nu}_j) \\& +
N_{\chiN}\Gamma(\chiN|\ell^+ _{i}W^-)\Br(W^-|  \bar q q ) - 
\tN_{\ell^+_i} \tN_{\bar q} \tN_{q} N_{\chiN}^{eq}\Gamma(l^+_{i} W^-|\chiN) 
\Br(W^-|  \bar q q) \\ & + 
N_{\chiN} \Gamma(\chiN|\ell^-_{j}W^+) \Br(W^+|\ell^+_i \nu_i) -
\tN_{\ell^-_{j}}\tN_{l^+_{i}}\tN_{\nu_i}N_{\chiN}^{eq} 
\Gamma(\ell^-_{j} W^+|\chiN) \Br(W^+|\ell^+_{i} \nu_i) \\& +
N_{\chiN} \Gamma(\chiN| Z \nu_j) 
\Br(Z |\ell^- _{i} \ell^+ _{i})  - 
\tN_{\nu_j}\tN_{\ell^-_i}\tN_{\ell^+_i} N_{\chiN}^{eq}  \Gamma (Z \bar{\nu}_j| \chiN)
\Br(Z |\ell^- _{i}\ell^+_{i}) \\ & + 
N_{\chiN} \Gamma(\chiN| Z \bar{\nu}_j) 
\Br(Z |\ell^- _{i} \ell^+ _{i})  - 
\tN_{\bar{\nu}j}\tN_{\ell^-_i}\tN_{\ell^+_i}  N_{\chiN}^{eq}  \Gamma (Z \bar{\nu}_j| \chiN)
\Br(Z |\ell^- _{i}\ell^+_{i})
\Bigr]\\ & +\hat{\sigma}(\chiN \chiN|\ell^+_i \ell^-_i) (N^2_{\chiN}-\tN_{\ell^+_i}\tN_{\ell^-_i} {N_{\chiN}^{eq}}^2) 
-\frac{\gamma(x)}{12}\Bigl[\delta_B+\eta(x)\delta_L\Bigr]
\end{split}
\end{align}
}

\end{appendix}

\end{document}